\documentclass[11pt]{article}

\usepackage[utf8]{inputenc}
\usepackage[T1]{fontenc}
\usepackage[margin=1in]{geometry}
\usepackage{times}
\usepackage{amsmath,amssymb}
\usepackage{booktabs}
\usepackage{array}
\usepackage{graphicx}
\usepackage{caption}
\usepackage{listings}
\usepackage{xcolor}
\usepackage{url}
\usepackage[hidelinks]{hyperref}
\usepackage{authblk}

\lstdefinestyle{tree}{
  basicstyle=\ttfamily\footnotesize,
  frame=single,
  framesep=6pt,
  rulecolor=\color{gray!50},
  columns=fullflexible,
  keepspaces=true,
  breaklines=false
}

\newcommand{\code}[1]{\texttt{\small #1}}

\graphicspath{{./}{figures/}{images/}}

\newcommand{\figplace}[2]{%
  \IfFileExists{#1}{\includegraphics[width=#2\textwidth]{#1}}{%
    \fbox{\parbox[c][4.5cm][c]{#2\textwidth}{\centering
      \textsf{\small [ figure placeholder ]}\\[4pt]
      \textsf{\footnotesize \detokenize{#1}}\\[4pt]
      \textsf{\scriptsize not found -- place this file beside the .tex}}}}}

\newcommand{\bibent}[1]{\noindent\hangindent=1.6em\hangafter=1 #1\par\vspace{0.45em}}

\title{\Large\textbf{Curriculum Brain: Constructing Curriculum Knowledge Graphs\\
as a Substrate for Cognitive Diagnosis}}

\author[1*]{Shrideep Tamboli}
\author[1]{Chiranjeevi Maddala}
\author[1]{Eshal Minhaj}
\affil[1]{AI Ready School}
\date{\small
*Corresponding author:
\textcolor{blue}{\href{mailto:27.tshrideep@gmail.com}{27.tshrideep@gmail.com}};\\
Contributing authors:
\textcolor{blue}{\href{mailto:chirans@gmail.com}{chirans@gmail.com}};
\textcolor{blue}{\href{mailto:eshalminhaj12@gmail.com}{eshalminhaj12@gmail.com}};}

\begin{document}
\maketitle

\begin{center}
\begin{minipage}{0.92\textwidth}
\small\raggedright
\textbf{Resources}\\[2pt]
Agents code: \url{https://github.com/MaximusTitan/q-matrix-agents}\\
Knowledge base template: \url{https://github.com/MaximusTitan/q-matrix-kb-template}\\
Dataset: \url{https://github.com/MaximusTitan/q-matrix-dataset}\\
Live curriculum visualization: \url{https://app.aireadyschool.com/curriculum}
\end{minipage}
\end{center}

\vspace{0.5em}
\hrule
\vspace{1em}

\begin{abstract}
Cognitive Diagnostic Models (CDMs) recover fine-grained, per-skill mastery
estimates rather than a single aggregate score, yet they remain largely unused
in practice. The obstacle is not the models but their precondition: the
Q-matrix, a mapping from every assessment item to the skills that item
requires. Q-matrices have historically been authored by hand through expert
judgment, and that construction step, not the diagnostic modeling, is the
binding constraint on deployment at scale.

We reframe the problem rather than shortcut it. Instead of building a fixed,
pre-mapped question bank, we separate the task into two stages: first construct
the curriculum's own knowledge graph, the full space of concepts and skills a
board's curriculum contains, independent of any particular item; then map items
against that graph on demand. This paper addresses the first stage only. The
item-mapping stage is designed but not implemented in the work reported here,
and the claim that separating the two shifts judgment cost from once per item to
once per curriculum is accordingly a design rationale that we set out and do not
test: no item is mapped anywhere in this paper, and we report no per-item
mapping cost.

We present \textbf{Curriculum Brain}, a two-repository system pairing a
persistent, version-controlled knowledge base with an agentic pipeline of eleven
single-responsibility agents coordinated by a thin deterministic orchestrator.
The system generates candidate concept-skill mappings from official curriculum
documents, checks them against accumulated rules, and compares them with a
concept-skill map extracted independently from the chapter's textbook, reporting
what the candidate adds and what it misses. In the automated configuration
evaluated here, that comparison is also enforced: the system repairs its own
failures and escalates to a human only when it cannot resolve a case itself. A
report-only mode instead leaves the textbook comparison to human judgment. The
system also implements a rejection path that encodes a human's
stated reason as a standing rule enforced on all future runs for that grade,
though that path was not exercised during the period reported here. Across 241
chapter runs (168 distinct chapters) through the full verification pipeline,
41.5\% of runs produced a Generator output that passed both checks without a
Doctor patch, almost always on a later attempt after feedback naming what was
missing, and 67.6\% resolved without escalation. Both figures are measured against criteria the system itself
produced, the accumulated ruleset for Check~1 and the extracted concept-skill map
for Check~2, so both describe internal consistency rather than agreement with an
external standard; and both pool two pipeline configurations, separated by a single change at run~77
that altered the verification gate, the models, the attempt budget and the retry loop
together; after it, the same figures are 57.0\% and 91.5\%. A scoped ablation finds
that the Generator's model upgrade does not account for the difference. All but two of the 168 chapters, both outside the release, were eventually
confirmed, though
confirmation is the loop's own termination condition rather than an independent
outcome. Total observed spend was \$216.77, or \$1.19 per chapter for the 154
chapters processed end to end through the current pipeline; averaged across all
223 released chapters the figure is \$0.97, diluted by 69 released chapters with
no full-mode run record. All costs are API spend only and exclude human review
time. We release both the framework and the resulting curriculum dataset.
\end{abstract}


\vspace{1em}

\section{Introduction}
\label{sec:intro}

CDMs can tell you exactly which skills a student has and hasn't mastered (a much
richer signal than a single test score or grade), yet they are barely used in
practice. The reason is not the models. It is what has to exist before the model
runs. This is the Q-matrix: the mapping of every question to the skills or
assessment objectives it requires. Every CDM depends on one. In practice, this
means manually categorizing questions from a pre-defined question bank into
assessment objectives: reading curriculum documents (published by the education
board, textbooks, etc.), extracting skills (assessment objectives), mapping each
question to the assessment objective(s) it targets, cross-checking that mapping
against what is actually taught, and tracing prerequisites within and across
chapters, grades, and subjects, all without an answer key, relying purely on
expert judgment applied over and over. Repeating this across 500+ chapters, for
every subject and grade, is what makes the approach impractical at scale. And
even where it succeeds, the resulting Q-matrix is fixed to whatever questions the
bank already contains: a skill can only be assessed if a question for it already
exists, and mapping a question from any new source means repeating the same
manual process all over again.

Our approach starts from a reframe of the problem rather than a shortcut through
it. Instead of treating Q-matrix construction as building a fixed, pre-mapped
question bank, we split the task in two: first construct the curriculum's own
knowledge graph (the full space of concepts and skills a board's curriculum
actually contains, independent of any specific question) and then map any
question, from any source, against that graph on demand. This paper covers the
first half only. The mapping stage is designed but not implemented in the work
reported here: we do not demonstrate it on a single item, and we do not measure
what mapping an item against a completed graph costs. We expect the separation to
be what makes automation tractable, moving the expensive, judgment-heavy work to
once per curriculum rather than once per question, but on the evidence presented
here that remains a design rationale rather than a finding.

This paper's contribution is \textbf{Curriculum Brain}: a two-repository system
(a knowledge base and a multi-agent workflow with human in the loop, working
together) that turns official curriculum documents into a prerequisite-linked
curriculum knowledge graph, with bounded human oversight rather than either full
automation or full manual effort. The verification loop described in
\S\ref{sec:method} covers the concept-skill layer of that graph; the prerequisite
edges do not pass through it (\S\ref{sec:discussion}), so verification applies to
one layer and not the other. We implement this as a multi-agent system (for a
broader survey of LLM agents in education, see Chu et al., 2025): eleven
single-responsibility agents, coordinated by a thin, deterministic orchestrator,
operating over a persistent, version-controlled knowledge base that accumulates
curriculum structure, generation rules, and human corrections across every run.
The system generates, evaluates, and repairs its own output, escalating to a
human only when it cannot resolve a case on its own. It is designed to treat that
escalation, and every correction that follows it, as new information retained
permanently rather than a one-time fix. We report in \S\ref{sec:selection} and
\S\ref{sec:discussion} that this retention path, although implemented end to end,
did not fire during the runs reported here.

As a practical outcome of this work, we release the framework's code and the
curriculum dataset it has produced, so that other researchers and practitioners
building CDMs do not face the same manual bottleneck we set out to remove.

Throughout, we are explicit that this system is \textbf{semi-automated, not fully
automated}: a human verification step remains part of the pipeline by design. We
do not claim that this step establishes trustworthiness. \S\ref{sec:discussion}
reports what the verification actually consisted of, which is a single reviewer
who is also the first author, with the analysis of genuinely failed escalations
delegated to an LLM agent rather than performed by hand.

\section{Background \& Related Work}
\label{sec:background}

\subsection{Cognitive Diagnosis and the Q-Matrix Bottleneck}
\label{sec:cdm}

Cognitive Diagnostic Models estimate a student's mastery across a set of latent
skills or attributes, rather than collapsing performance into a single score.
Every such model depends on a Q-matrix (a mapping between assessment items and
the skills each item requires), first formalized in the rule-space tradition
(Tatsuoka, 1983) and extended through conjunctive and disjunctive model families
such as DINA (Junker \& Sijtsma, 2001), DINO (Templin \& Henson, 2006), and
G-DINA (de la Torre, 2011). A related psychometric tradition, Knowledge Space
Theory (Doignon \& Falmagne, 1985), models mastery as partially ordered knowledge
states rather than a fixed item-skill matrix; it informs the same broader goal of
fine-grained diagnosis without sharing the Q-matrix formalism itself. The
diagnostic value of these models is well established; their deployment is not.
The Q-matrix has historically been built by hand, through expert judgment or
verbal report analysis, and this construction step (not the diagnostic modeling
itself) is the actual barrier to using CDMs at scale.

\subsection{Automating Q-Matrix Construction}
\label{sec:autoq}

A growing body of work uses LLMs to automate Q-matrix construction, though nearly
all of it targets item-level mapping against existing instruments or response
data rather than curriculum-level construction from source documents. Zhang
et al. (2026a) generate candidate item--misconception mappings via
chain-of-thought prompting and validate them with NeuralCDM (a neural CDM
treating the Q-matrix as a learnable parameter): on a 26-item thermodynamics
instrument ($\sim$600 responses), their best configuration exceeded the expert
baseline's model fit. Other approaches construct Q-matrices directly from item
text and attribute definitions with generative models, without an initial matrix,
validated psychometrically against expert- and student-derived baselines (Du et
al., 2026); encode concept descriptions with a pretrained language model to
correct errors in an initial Q-matrix while predicting student performance (Fan,
J. et al., 2026); use LLM contextual prompting to construct Q-matrices that then
constrain adversarial data augmentation (Zhu et al., 2026); or embed a given Q-matrix as a structural constraint in a neural network
that diagnoses skill mastery from response data rather than from curriculum
content (Tao et al., 2024). A smaller study scores AI-generated Q-matrices
against an expert-validated one directly, by inter-rater agreement rather than by
model fit, finding wide variation across models on a 20-item reading test: the
strongest reached $\kappa = 0.63$ against the validated matrix, above all four
human raters, though the remaining models fell below them, and a repeat a year
later on newer model versions scored lower still (Fan, K. et al., 2026). A related line uses expert-supervised LLM annotation to build
Q-matrices over benchmark items whose examinees are LLMs rather than students,
estimating fine-grained model abilities rather than constructing a curriculum
(Zhang et al., 2026b).

This confirms LLM-driven Q-matrix construction is an active, multi-group
direction. What distinguishes our approach is its unit and source: existing work
maps items against a fixed instrument or infers structure from response data,
whereas we build the curriculum's own knowledge graph from official curriculum
documents first, independent of any item, and map items against it afterward.

\subsection{Curriculum Knowledge Graph Construction}
\label{sec:ckg}

Closest to our system's architecture is K12-KGraph (Liang et al., 2026), a
curriculum-aligned knowledge graph extracted from official Chinese K--12
textbooks via a five-stage pipeline: OCR-based document parsing, section
segmentation, LLM schema-guided extraction of nodes and edges, hierarchical
bottom-up merging, and DAG validation on taxonomic and prerequisite subgraphs,
followed by human verification. Both systems share a core premise: curriculum
structure extracted at scale via LLMs and checked by humans is a viable
foundation for downstream educational tasks.

A more fundamental difference lies in the role textbooks play in each pipeline.
K12-KGraph generates its knowledge graph directly from textbook content; in our
system, the Generator never reads the textbook, and the textbook's role is
comparison rather than construction. That separation is complete in the
framework's report-only mode, where textbook coverage is shown to a human and
never fed back into generation. It is partial in the automated gate mode used for
every run reported here: from a candidate's second attempt onward, the
textbook-derived items it is missing are passed back to the Generator by name, so
the textbook does contribute content, and \S\ref{sec:repairmech} measures how
much. We treat textbook text as noisy for graph construction: it is full
of terms that resemble a chapter's core concepts in wording but are not concepts
the curriculum actually targets, and an LLM extracting directly from textbook
prose readily mistakes these near-synonyms for real concepts, inflating the graph
with pseudo-concepts. Two qualifications belong here, because we do not avoid
textbooks entirely. First, our own extraction applies controls against precisely
this failure: it excludes structural and review headings by name, and it merges
any two candidate concepts a student would answer with the same explanation
(\S\ref{sec:start}). Second, we retain the textbook downstream as the reference
the generated output is compared with (\S\ref{sec:verify}), and in gate mode as a
criterion it must meet, because it is the only source in the verification loop
independent of the curriculum documents the Generator reads; using curriculum
documents on both sides would make the check a test of the Generator against its
own input. In gate mode that independence holds for a candidate's first attempt
only, for the reason given in \S\ref{sec:verify}. What that choice
costs, and what it does to the released data's provenance, is measured in
\S\ref{sec:repairmech} and discussed in \S\ref{sec:discussion}. This tracks the
curriculum-alignment literature, where textbook-derived concept sets have been
found several times larger than the corresponding curriculum specification (Kim
et al., 2022), and textbook editions have been found to align more closely with
one another than with the standard itself (Yu et al., 2022). We instead build the graph from the curriculum documentation
education boards publish publicly, which already states the exact learning and
assessment objectives by chapter, grade, and subject, keeping the graph anchored
to what the curriculum requires rather than to a textbook's incidental
vocabulary.

The two systems also differ in purpose and in how human verification is used.
K12-KGraph's downstream use is benchmark construction and instruction-tuning data
synthesis for educational LLMs, not Q-matrix generation for cognitive diagnosis,
and its human verification functions as a one-time QA pass whose corrections do
not appear to persist as rules. In our system, a human correction is encoded as a
standing rule enforced automatically on all future runs (\S\ref{sec:method}), so
the intended difference is not human-in-the-loop versus not (both systems have
it) but whether the loop changes the system going forward or only the artifact in
front of it. We offer this as a design contrast rather than a demonstrated one:
no such rule was created during the period reported here
(\S\ref{sec:selection}, \S\ref{sec:discussion}). A related but
differently-purposed system, MDKAG (Zhao et al., 2025a), builds a multimodal
disciplinary knowledge graph from textbooks, lecture slides, and classroom video
to power retrieval-augmented educational QA rather than curriculum-level
Q-matrix construction, illustrating the same underlying resource, a structured
knowledge graph over educational content, applied to a different downstream task.

\subsection{Self-Improving Prompts in Agentic Pipelines}
\label{sec:selfimprove}

Our self-tuning prompt mechanism draws on prior work in automatic prompt
optimization: LLMs generating and selecting among candidate prompts against a
scoring function (Zhou et al., 2022; Yang et al., 2023), treating prompt failure
as a signal to rewrite rather than a fixed error (Pryzant et al., 2023),
iterative self-feedback loops that revise output against explicit criteria
(Madaan et al., 2023), and prompts mutated based on grounded feedback from an
agent's own environment or execution outcomes (Fernandes et al., 2026). Nearest
to our rule-accumulation mechanism specifically is Pei et al.'s (2025) synthesis
of standing guidelines from execution traces, a similar move from correcting
individual outputs to accumulating durable, reusable rules, though theirs are
synthesized automatically rather than authored by a human reviewer as ours are.

For the overall comparison developed in the rest of this section, though, closest
to our mechanism is Zhang et al.'s (2026a) Lightweight Expert-in-the-Loop
protocol, in which the prompt evolves through four successive versions, from a
zero-shot baseline through a structured misconception library and explicit
evaluation criteria to few-shot examples drawn from expert annotations, with
NeuralCDM model fit improving from the first structured version to the
expert-informed one. Their framing
shifts expert effort upstream, from correcting individual outputs to calibrating
the prompt that generates them, so expert effort no longer scales linearly with
item-bank size.

Our approach differs in who performs the rewriting and how the result persists.
Zhang et al.'s (2026a) four prompt versions are manually authored by the research
team over the study, with expert annotations entering at the final version through
human-driven iteration. In
our system, two separate mechanisms are designed to divide this work: the
Revision agent automatically rewrites the prompt in response to a verification
failure, but only for the current run, never persisted (since run~77; earlier
rewrites were written back, \S\ref{sec:escalation}); and a human explicitly
rejecting a chapter's output appends the rejection reason to a standing rule file
for that grade and subject, loaded automatically on every future run. The
intended distinction is not that expert-guided prompt calibration is novel to us
(it plainly is not) but that persisting a correction should not require a
researcher to manually re-version a prompt file: a single rejection with a reason
would be enough, and it is the human's judgment of what went wrong, not their
hand at rewriting a prompt, that becomes the standing rule.

In practice the second mechanism was not used: no grade-scoped rule file was created
at any point in the run history reported here. The first mechanism, for most of the
period before run~77, did not behave as described above. Until then the Revision
agent's rewrites were written back to the shared prompt library, so the Mathematics
base prompt and five grade-level overrides accumulated revisions from individual
chapters' attempt loops, including chapter-specific material such as ``Egyptian
fractions'' carried into a subject-wide prompt. At run~77 that write-back was
removed, the grade overrides were deleted, and the Mathematics base prompt was pruned
by hand from 1,544 to 356 words (\S\ref{sec:escalation}); the Science base prompt,
itself produced by an early Revision rewrite, remained the seed for every reported
Science run. On the evidence of this study, then, our position relative to Zhang et
al. (2026a) is a difference in available mechanism rather than in demonstrated
practice: automatic prompt persistence did occur, produced the chapter-specific
leakage \S\ref{sec:start} anticipates, and was replaced by human curation.

\section{System Architecture}
\label{sec:arch}

Curriculum Brain is built as two repositories with a clean separation of
responsibility: a knowledge base that owns all persistent state, and an agentic
pipeline that acts on it. Neither is meaningful in isolation: the knowledge base
has no logic of its own, and the agents retain no state without it. This
separation is deliberate: curriculum knowledge (what's been extracted, what rules
have been learned, what's been confirmed) can be inspected, versioned, and
audited independently of any change to the code that produces it.

The two repositories operate over a shared data model, though at different levels
of it. In the agentic pipeline, each output row identifies a concept-skill pair by
board, subject, grade, chapter, concept, and skill: a flat identifier hierarchy,
not a node hierarchy. The knowledge graph itself has only one node type, the
concept; board, subject, grade, and chapter are properties on that node, and
skills are reduced further, to a per-concept count backed by a separate skill list
rather than their own graph nodes or edges: skill-level prerequisite data still
exists in the underlying CSVs (\S\ref{sec:prereq}, \S\ref{sec:schema}) and drives
some of the concept-level edges shown here (\S\ref{sec:viz}), it just never
surfaces as its own graph entity. Prerequisite relationships are typed edges
between concept nodes at four levels: within a chapter (L1), across chapters of
the same grade and subject (L2), across grades within the same subject (L3), and,
planned but not yet implemented, across subjects (L4).

\subsection{Knowledge Base}
\label{sec:kb}

The knowledge base is the system's single source of truth. It is a curriculum
content and QA repository, not a codebase: no application logic lives here, only
the data the agents read from and write to.

\begin{lstlisting}[style=tree]
q-matrix-kb/
|-- rulesets/            # universal_rules.md: the one manually authored input
|-- textbooks/CBSE/      # Source textbook PDFs, Subject -> Grade -> Chapter
|-- curriculum-docs/CBSE/  # Official curriculum/teacher docs, Subject -> Grade
|-- prompt-library/CBSE/   # System-authored base_prompt.md per subject
`-- escalations/CBSE/      # Per-chapter, per-run generation history
    `-- Subject/Grade/Chapter/<date>/
        |-- attempt_N_prompt.md
        |-- gen_attempt_N.csv
        |-- doctored_attempt_N.csv
        |-- doctored_rules_attempt_N.csv
        |-- last_csv.csv
        |-- report.md
        `-- run.json
\end{lstlisting}

Each subject/grade/chapter combination has a matching folder across
\code{textbooks/}, \code{curriculum-docs/}, and \code{escalations/}, so the source
material, the official curriculum reference, and the full generation history for a
given chapter are always co-located. The escalation folders in particular double
as an audit trail: every attempt's prompt, the generated CSV, any surgically
doctored variant, and the structured run record are retained per date, not
overwritten, so an escalated chapter's entire generation history remains
inspectable after the fact, not just its final confirmed output. This retention is
specific to escalation: a run that resolves on its own writes no escalation
folder, and the resolved prompt behind each of the 163 confirmed full-pipeline runs
is recoverable only from the knowledge base's git history, not from its working
tree. Prompt retention is therefore nearly complete for failures (the escalation
folders of two superseded Grade~4 chapters survive only in git history) and partial
for successes, the reverse of what a reproducibility audit would want.

Every pipeline run pulls the latest state of this repository before starting and
pushes any new or corrected state back to it on completion, so the knowledge base
reflects, at any point in time, everything the system has learned so far across
every chapter. The design supports several operators sharing one knowledge base,
though in the runs reported here there was only one (\S\ref{sec:discussion}).

\subsection{Agents}
\label{sec:agents}

The agentic pipeline is the system's code layer: a thin orchestrator with no
reasoning logic of its own, coordinating eleven single-responsibility agents,
backed by a FastAPI service and a live dashboard for monitoring runs. The
orchestrator's role is purely one of control flow: deciding which agent runs next
and under what condition, while every judgment call (what counts as a valid skill,
whether a candidate passes, which candidate is best) is made by an agent, not the
orchestrator itself.

Each agent is defined not just by its responsibility but by which shared skill
modules it draws on, a useful lens on the system's actual coupling, since it shows
which capabilities (LLM access, KB reads/writes, CSV validation) are load-bearing
across the pipeline versus specific to one agent's job.

\begin{table}[htbp]
\centering
\caption{The eleven agents, plus the orchestrator, and the shared skill modules
each depends on.}
\label{tab:agents}
\footnotesize
\begin{tabular}{@{}>{\raggedright\arraybackslash}p{0.16\textwidth}
                  >{\raggedright\arraybackslash}p{0.52\textwidth}
                  >{\raggedright\arraybackslash}p{0.26\textwidth}@{}}
\toprule
\textbf{Agent} & \textbf{Description} & \textbf{Skills used} \\
\midrule
Map Extraction & Reads the chapter's textbook PDF and extracts a concept-skill
map, used later only to compare against the Generator's output, not to produce it.
& \code{pdf\_reader}, \code{kb\_access}, \code{llm} \\
\addlinespace
Generator & Produces a candidate concept-skill CSV from the resolved prompt and
the curriculum documents.
& \code{kb\_access}, \code{llm}, \code{csv\_utils} \\
\addlinespace
Eval & Runs Check~1 (rule compliance) and Check~2 (comparison with the
concept-skill map) on a candidate in parallel, returning structured feedback
rather than a pass/fail flag. Check~2 is blocking in gate mode and report-only
otherwise. & \code{kb\_access}, \code{llm}, \code{csv\_utils}, \code{diff} \\
\addlinespace
Doctor & Gate mode only. Surgically patches a candidate that passed Check~1 but
failed Check~2, fixing only the coverage gaps. & \code{llm}, \code{csv\_utils} \\
\addlinespace
Rules Doctor & Surgically patches a candidate that passed Check~2 but failed
Check~1, fixing only the flagged rule violations.
& \code{llm}, \code{csv\_utils} \\
\addlinespace
Revision & Rewrites the generation prompt itself when a candidate fails both
checks (or a single-check repair also fails), for the next attempt only.
& \code{llm} \\
\addlinespace
Judge & Selects the best candidate when two or more have independently passed
within the same run. Reachable only in runs up to 15 July 2026; the current loop
stops at the first passing candidate, so Judge is not invoked. & \code{llm} \\
\addlinespace
Prerequisites (L1) & Maps a confirmed chapter's within-chapter prerequisite edges
automatically, with no separate trigger. & \code{llm}, \code{csv\_utils} \\
\addlinespace
Chapter Relevance & Cheap, recall-biased pre-filter that narrows the pool of
candidate chapters before the more expensive L2/L3 comparison. & \code{llm} \\
\addlinespace
Prerequisite (L2) & Maps cross-chapter prerequisites within the same grade and
subject; manually triggered once every chapter in scope has L1 in place.
& \code{llm} \\
\addlinespace
Prerequisite (L3) & Maps cross-grade prerequisites within the same subject;
manually triggered once every earlier grade has L1 in place. & \code{llm} \\
\midrule
Orchestrator & Thin control-flow layer deciding which agent runs next; holds no
judgment of its own. & \code{kb\_access}, \code{csv\_utils}, \code{git\_sync},
\code{run\_record}, \code{report\_render}, \code{llm} \\
\bottomrule
\end{tabular}
\end{table}

\code{llm}, a shared wrapper around every agent's LLM calls, is the only skill
every single agent depends on, which is the clearest evidence that these are
LLM-driven judgment calls throughout, not rule-based logic with LLM assistance at
the edges. \code{kb\_access} and \code{csv\_utils} are the next most load-bearing,
reflecting that most agents read from or write to the knowledge base and validate
against the fixed CSV schema.

\begin{table}[htbp]
\centering
\caption{Shared skill modules.}
\label{tab:skills}
\small
\begin{tabular}{@{}>{\raggedright\arraybackslash}p{0.18\textwidth}
                  >{\raggedright\arraybackslash}p{0.76\textwidth}@{}}
\toprule
\textbf{Skill} & \textbf{Description} \\
\midrule
\code{llm} & Shared wrapper around the Vercel AI Gateway (OpenAI-compatible)
endpoint that every agent calls through. \\
\code{kb\_access} & Sole owner of KB folder-structure knowledge; all knowledge
base read/write operations. \\
\code{csv\_utils} & Parses and validates the fixed-schema curriculum CSVs produced
by the Generator Agent. \\
\code{diff} & Semantically diffs a generated CSV's concepts/skills against the
concept-skill-map for Eval Check~2. \\
\code{pdf\_reader} & Extracts text from PDF files (textbook chapters, curriculum
docs) via pdfplumber. \\
\code{git\_sync} & Pulls/pushes the KB repo's git state at the start/end of an
orchestrator run. \\
\code{run\_record} & Assembles the structured per-chapter run.json record from a
pipeline run's events. \\
\code{report\_render} & Renders the human-readable report.md from a structured run
record. \\
\code{file\_io} & Base, path-agnostic file read/write/exists operations that other
KB skills build on. \\
\code{model\_stats} & Aggregates persisted run records into per-(agent, model)
pass/fail and cost stats for the dashboard. \\
\bottomrule
\end{tabular}
\end{table}

\code{file\_io} and \code{model\_stats} aren't directly imported by any agent or
the orchestrator; \code{file\_io} is used by \code{kb\_access} and
\code{model\_stats}, and \code{model\_stats} by \code{api.py} and the tests. This isolated-failure-mode
design, where each agent's role is narrow enough that one agent's error doesn't
cascade into another's, mirrors other specialized multi-agent generation pipelines
such as OpenGloss (Bommarito, 2025), though ours adds the accumulated-rule and
repair-loop machinery described in \S\ref{sec:method}.

The specific responsibilities of each agent, and how they interact over the course
of processing a chapter, are described in \S\ref{sec:method}.

\subsection{Downstream Visualization}
\label{sec:viz}

A third, separate repository consumes the knowledge base's confirmed CSVs to
render an explorable 3D visualization of the curriculum knowledge graph. Unlike
the two core repositories, this component makes no LLM calls and no judgment of
its own: its export step is a deterministic, purely mechanical transformation of
confirmed KB data into a graph representation (nodes, edges, and, where a
prerequisite relationship was never directly evaluated between two concepts but
can be inferred from an already-judged skill-level relationship, a small number of
mechanically derived edges, clearly flagged as such in the output). This tool is
not part of the core generation/verification pipeline described in
\S\ref{sec:method}; it is a downstream consumer of the pipeline's output, worth
noting here as part of the released framework but discussed further as a
limitation-relevant detail in \S\ref{sec:discussion}.

\section{Methodology}
\label{sec:method}

At a high level, every chapter follows the same path. The system starts from a
chapter's curriculum documents and produces a candidate CSV mapping concepts to
skills. That candidate is checked against two things: whether it follows the
accumulated rules for that board, subject, and grade, and how it compares with
the concepts and skills the chapter's textbook contains. The second comparison
runs in one of two modes. In gate mode, used for every run reported in
\S\ref{sec:results}, it is enforced like the first; in report-only mode, it is
written to the run report for a human to act on and plays no part in the loop
(\S\ref{sec:verify}). Except where noted, this section describes gate mode. If
the candidate fails a check, the system repairs it (a small, targeted patch if only one check failed, or
a full rewrite of the generation prompt if both failed) and tries again. This
repeat-and-repair cycle continues until a candidate passes or the system judges
that further attempts will not help, at which point it hands off to a human
instead of continuing to guess. Once a chapter's mapping is confirmed, the system
maps its internal prerequisites automatically (L1), and, on request, its
prerequisites against other chapters in the same grade and subject (L2) and
against chapters in earlier grades within the same subject (L3). A fourth level,
cross-subject prerequisites (L4), is planned but not yet implemented.

\begin{figure}[htbp]
\centering
\figplace{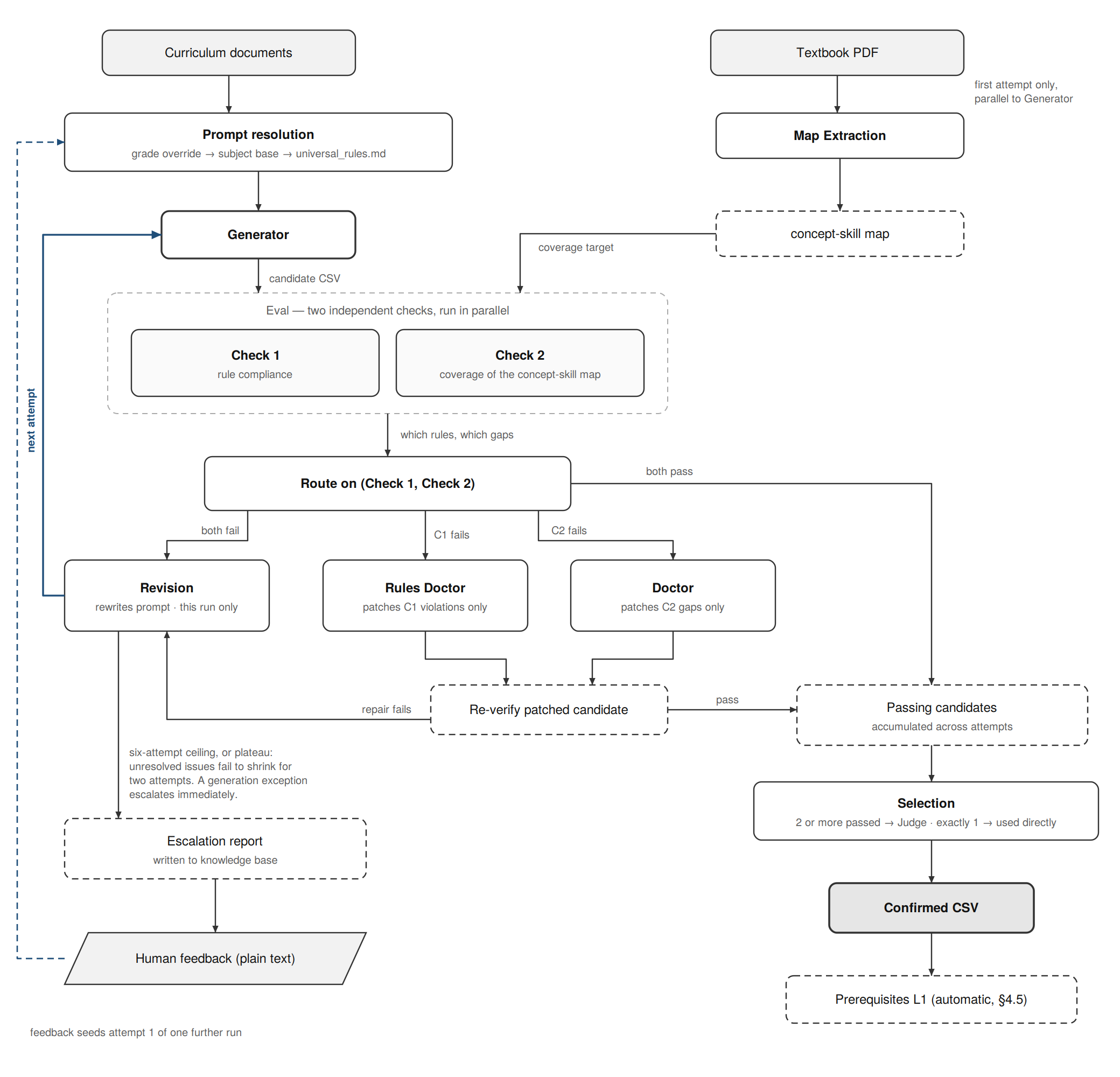}{0.95}
\caption{The full pipeline for a chapter: prompt resolution and the parallel Map
Extraction branch that produces the concept-skill map, through generation and
verification (Eval Check~1 and Check~2), the four routing outcomes with their
repair or revision paths, and resolution either into a confirmed CSV (through
Judge selection when several candidates had passed, a path the current loop no
longer reaches) or through escalation to a human, whose feedback seeds the first
attempt of one further run. Gate mode is shown. In report-only mode, Check~2's
output goes to the run report and does not feed routing, repair or revision.}
\label{fig:loop}
\end{figure}

That's the shape of it (Figure~\ref{fig:loop}). The rest of this section walks
through each phase in detail.

\subsection{Starting a Chapter}
\label{sec:start}

Before any agent produces anything, the orchestrator pulls the knowledge base's
latest state and checks what already exists for this chapter and subject. If the
chapter's concept-skill map does not exist yet, Map Extraction reads the chapter's
textbook PDF and produces one: the full set of concepts and skills the textbook
actually contains. This map is not what the Generator works from (that's the
curriculum documents, for the noisy-textbook reasons given in \S\ref{sec:ckg}); it
is used later, by the Eval agent's coverage check, to show where the Generator's
curriculum-doc-derived output differs from what the textbook covers. Because
the Generator never depends on Map Extraction's output at generation time, the two
run in parallel on a chapter's first attempt. Once written, the map is not revised
by the system: later attempts load the existing file rather than re-extracting, so
in gate mode a chapter's acceptance criterion is fixed for the whole of its
generation history. That is deliberate, since a criterion moving between attempts
would make pass and fail incomparable across them, but it also means that in gate
mode an extraction error is not correctable by any automated path; in report-only
mode it is one more item a human can decline (\S\ref{sec:discussion}).

The prompt the Generator runs against is resolved to whichever is most specific
and already exists: a grade-level override, failing that the subject's base
prompt, and failing that, \code{universal\_rules.md}, the system's single manually
authored input. Base and grade prompts are otherwise produced by the system; the one
exception in the reported runs is the Mathematics base prompt, pruned by hand at
run~77 (\S\ref{sec:escalation}). The very first time a subject is run, neither a base prompt nor a
grade override exists yet, so the Generator runs directly against the universal
ruleset, logged as a cold start. If a candidate fails either check, the Revision
agent rewrites the prompt through the same repair mechanism described in
\S\ref{sec:repair}, but only in memory, for that run. This is deliberate:
persisting a chapter-specific revision back to \code{base\_prompt.md} would risk
overfitting the shared prompt to that one chapter's quirks rather than
generalizing across the subject. The revised prompt is therefore never written
back: every chapter always starts its own attempts from the same clean,
unmodified seed, and one chapter's revisions can never carry over into another's.
This describes the pipeline from run~77 onward; before that boundary, rewrites were
written back to the shared prompt library (\S\ref{sec:escalation}).

\subsection{Generating and Verifying a Candidate}
\label{sec:verify}

The Generator takes the resolved prompt and the curriculum documents and produces
a candidate CSV; the concept-skill map plays no role here; it is consumed later,
by Eval's coverage check. Every candidate is then checked by Eval against two
independent criteria, run in parallel:

\begin{itemize}
  \item \textbf{Check 1 (rule compliance):} does the candidate follow the
  accumulated ruleset for this board, subject, and grade?
  \item \textbf{Check 2 (coverage):} how does the candidate compare with the
  concept-skill map? It returns two lists: items in the map the candidate lacks
  (missing), and items in the candidate that correspond to nothing in the map
  (extra).
\end{itemize}

Check~2 is designed as a review aid. A practitioner building a Q-matrix can read
both lists and decide, item by item, whether a missing textbook concept belongs in
the curriculum mapping and whether an extra one should stay; the textbook is
evidence for that judgment, not an authority over it. The framework supports this
directly in report-only mode (\code{-{}-check2-mode report}). Check~2 then runs
once on the final candidate and its two lists go into the run report; it does not
fail the run, does not invoke the Doctor, and its lists are not fed back into
generation. The practitioner edits the CSV and supplies it for prerequisite
mapping (\S\ref{sec:selection}).

Gate mode (\code{-{}-check2-mode gate}, the default) instead resolves the
comparison automatically, so that the pipeline can run end to end without
per-chapter human review. We added it for throughput, and every run reported in
\S\ref{sec:results} uses it. It treats the two lists asymmetrically. A missing
item is a blocking failure: the repair agent is instructed that all such items must
be covered by its edit, and on any further attempt the missing items are passed to
the Generator by name in its retry feedback. An extra item is not a failure: it is
reported, and the repair agent decides per item whether to keep it as valid chapter
content or drop it as noise, under an instruction to be conservative and remove
only clear non-content. Gate mode is therefore recall-oriented, on the view that
for a graph intended as a substrate a spurious skill can be ignored downstream
whereas a missing one silently bounds what can ever be assessed. Two consequences
follow. From a candidate's second attempt onward the Generator has been shown what
Check~2 expects, so only first-attempt coverage measures its unaided agreement with
the textbook. And gate mode trades curation for automation;
\S\ref{sec:repairmech} reports what that costs in practice.

Eval's output is structured, not just pass/fail: Check~1 returns the specific rule
violations, Check~2 returns the specific missing concepts and skills.

Check~1's implementation changed partway through the run history reported in
\S\ref{sec:results}, and the two versions are not the same test. In the earlier
configuration a single LLM call judged the candidate against the full 31-rule
ruleset. In the current one, structural and identifier-fidelity rules are verified
deterministically in code, the LLM judges only content-quality rules, of which
eight are blocking, and three further rules return non-blocking advisory flags
rather than failures. \S\ref{sec:passrates} therefore reports results for the two
periods separately, and \S\ref{sec:escalation} describes the change and its
consequences.

\subsection{Repair, Revision, and the Attempt Loop}
\label{sec:repair}

What happens next is determined entirely by which of the two checks failed
(Table~\ref{tab:routing}). Both single-check cases route to a repair agent that
touches only the failing dimension and leaves the satisfied one alone. Both checks
failing at once routes to Revision instead, because Doctor and Rules Doctor exist
specifically to fix one isolated problem, not two simultaneously; a single-check
repair that then fails re-verification falls through to Revision on the same
reasoning. The escalation triggers sit outside this grid and are described at the
end of this section.

\begin{table}[htbp]
\centering
\caption{Routing on the (Check 1, Check 2) outcome pair.}
\label{tab:routing}
\small
\begin{tabular}{@{}llp{0.18\textwidth}>{\raggedright\arraybackslash}p{0.42\textwidth}@{}}
\toprule
\textbf{Check 1 (rules)} & \textbf{Check 2 (coverage)} & \textbf{Next step} &
\textbf{What the step changes} \\
\midrule
Pass & Pass & Selection (\S\ref{sec:selection}) & Nothing; the candidate joins the
run's pool of passing candidates \\
\addlinespace
Pass & Fail & Doctor & Patches only the missing concepts and skills Eval listed;
rows that already satisfy the rules are left untouched \\
\addlinespace
Fail & Pass & Rules Doctor & Patches only the rule violations Eval flagged;
coverage is left untouched \\
\addlinespace
Fail & Fail & Revision & Rewrites the generation prompt for the current run; the
Generator reruns against the new prompt \\
\bottomrule
\end{tabular}
\end{table}

In report-only mode Check~2 does not take part in routing: each attempt is
evaluated by Check~1 alone, and Check~2 is treated as passing, so only the Check~1
column of Table~\ref{tab:routing} applies. A candidate that passes Check~1 is
confirmed. A Check~1 failure goes to Rules Doctor, which in this mode does not
receive the concept-skill map, and to Revision if that repair fails
re-verification; Revision then receives Check~1 feedback only. Check~2 runs once,
on the confirmed candidate, before prerequisite mapping. A run that escalates
without any candidate passing Check~1 never runs it.

Revision is the one agent that changes what the chapter's next attempts will
produce, not just the candidate in front of it, though as established in
\S\ref{sec:start}, that change never outlives the current run. Given the failure
history and which check or checks failed, it rewrites the prompt itself: a new
subject-level base prompt if the subject has no working prompt yet, or a
grade-specific override if the base prompt already works everywhere else but keeps
failing for this one grade. The Generator then reruns against the new prompt, and
the check-and-repair cycle begins again.

This continues, attempt after attempt, until a candidate passes or the system
decides continuing will not help. That decision rests on two conditions, whichever
triggers first: a hard ceiling on attempts, or a plateau, the total count of
unresolved issues (summing Check~1's rule violations and Check~2's missing concepts
and skills in gate mode, and Check~1's violations alone in report-only mode)
failing to shrink for two attempts in a row. The ceiling is six
attempts in the system as described here, but it was three for the first 76 of the
runs reported in \S\ref{sec:results}, raised at the same point as the Check~1
change above. The plateau condition was introduced at the same point
(\S\ref{sec:escalation}), so before it a run that kept failing continued to the
ceiling. No run before that point reached a fourth attempt, which is why
\S\ref{sec:passrates} reports the attempt distribution for the two periods
separately. A separate, unconditional trigger also exists for outright generation
failures: an exception during generation itself escalates immediately, regardless
of attempt count or plateau state.

\subsection{Selection and Human-in-the-Loop}
\label{sec:selection}

When the loop resolves on its own, the passing candidate becomes the chapter's
confirmed CSV. It can come from the Generator directly, a Doctor-patched fix, or a
Rules-Doctor-patched fix. Earlier in the run history the loop continued after a
pass, so several passing candidates could accumulate across attempts, and Judge
selected the best of them. This happened in 6 runs, all on or before 15 July 2026
(UTC); one of them predates the current run-record schema and records no Judge
model or cost, which is why Table~\ref{tab:cost} counts five. Since then the loop
stops at the first passing candidate, so Judge is retained in the code but not
invoked.

In report-only mode the human's role moves from resolving escalations to reviewing
coverage. Once a candidate passes Check~1, the run report lists the textbook items
it is missing and the items it contains that the textbook does not. The
practitioner adds or removes items by hand, treats the edited file as the chapter's
final CSV, and runs within-chapter prerequisite mapping on it directly
(\code{-{}-prereq-csv}). This is the workflow the design intends for users building
a Q-matrix they will curate themselves. It has three limits. The edited file passes
through neither Check~1 nor Check~2, so the practitioner's changes are not checked
against the ruleset. Only L1 is regenerated, and any existing L2 and L3 columns are
dropped, so cross-chapter and cross-grade mapping must be re-run afterwards. And the
chapter the file is written to is taken from the file's own board, subject, grade
and chapter columns, not chosen separately.

When the loop does not resolve on its own, the orchestrator writes a structured
escalation report to the knowledge base. A human reads it and resubmits the
chapter with feedback as plain text; that feedback is injected into the very first
attempt of the new run only, after which the pipeline proceeds through its normal
repair loop, up to the same attempt ceiling in force at the time
(\S\ref{sec:repair}), without further human input. That is the mechanism the
system provides. In the runs reported here it was exercised differently, and
\S\ref{sec:discussion} gives the detail: most resubmissions followed a
configuration change rather than a per-chapter reading, and for the chapters that
failed under the stable configuration the escalation report was read and the
resubmission feedback written by an LLM coding agent rather than by the author.

Separately, even a chapter that passed entirely on its own can later be rejected by
a human with a stated reason; that reason is written directly into the ruleset for
the relevant board, subject, and grade, and Eval enforces it automatically on every
future run for that grade, with nothing needing to be re-authored. This rejection
path is implemented end to end, from the command line and the service API through
to the ruleset file, but no chapter was rejected through it during the runs
reported here. No grade-scoped ruleset therefore exists, and the accumulated
ruleset Check~1 enforced throughout is the single manually authored
\code{universal\_rules.md} described in \S\ref{sec:start}. Neither feedback path
requires a human to touch a prompt or a line of code directly; both are designed to
feed back as plain text that the orchestrator encodes into the knowledge base,
which is what would let the system's judgment keep improving run over run without
depending on continuous human review of its output. We describe what the mechanism
does, not an observed effect of it. The resubmission path was used, for 71 of the
168 chapters, but the feedback text supplied at resubmission is not itself
persisted to the knowledge base, so those interventions appear in the run records
only as a subsequent run and not as their content.

\subsection{Prerequisite Mapping}
\label{sec:prereq}

Once a chapter's CSV is confirmed, prerequisite mapping proceeds through up to
three implemented levels, only the first of which runs automatically:

\begin{itemize}
  \item \textbf{L1 (within-chapter, automatic):} maps a chapter's internal
  prerequisite edges, concept-to-concept and skill-to-skill, immediately after
  selection, with no separate trigger.
  \item \textbf{L2 (cross-chapter, same grade and subject, manually triggered):}
  runs only once every chapter in that grade and subject already has L1 in place.
  This gate checks the confirmed CSV for non-empty L1 data, not whether the tracked
  Prerequisites (L1) agent specifically produced it, so a chapter confirmed through
  an earlier, untracked pathway can still satisfy it, which is why L2 ultimately
  covers more chapters than the automatically-tracked L1 population
  (\S\ref{sec:cost}). Chapter Relevance pre-filters the pool of sibling chapters
  first, a cheap, recall-biased pass that narrows which chapters are worth the more
  expensive full comparison.
  \item \textbf{L3 (cross-grade, same subject, manually triggered):} extends
  prerequisite edges across grade boundaries, gated on every earlier grade in the
  subject already having L1 in place, and reuses the same Chapter Relevance
  pre-filter as L2, run once per earlier grade.
\end{itemize}

Verification across all three levels is a simple two-attempt parse retry, not the
Eval-driven Check~1/Check~2/repair/escalation loop that governs the main
per-chapter pipeline (\S\ref{sec:verify}--\S\ref{sec:selection}); a failure falls
back to empty prerequisite columns rather than escalating to a human. This
asymmetry is a design decision rather than an omission, and it follows from the two
tasks having different shapes. Concept-skill generation must satisfy an accumulated
ruleset that standardises what counts as a concept and as a well-formed skill, and
it can fail in several independent ways at once: wrong granularity, invented
content, missing coverage, malformed rows. A check-and-repair loop is warranted
there because there are many ways to be wrong and most of them are correctable in
place. Prerequisite mapping is a constrained pairwise judgment made over two
concept lists that are supplied to it, with few rules to satisfy beyond returning a
parseable answer, so its characteristic failure is a parse failure rather than a
rule violation, which is what the retry addresses. Model selection follows the same
reasoning: L1 compares concepts within a single chapter, a small and semantically
coherent context, and ran predominantly on the cheaper model, whereas L2 and L3
must reason across a whole grade or across earlier grades and ran on the stronger
one (\S\ref{sec:repro}, \S\ref{sec:cost}). What this design does not provide is
stated in \S\ref{sec:discussion}: the prerequisite layer carries a well-formedness
check and no correctness check. Cross-subject mapping (L4) is anticipated in the
system's design but not yet implemented (no module, prompt, endpoint, or model key
exists for it), making L3 the current ceiling of what the pipeline actually
produces.

\section{Releasing the Framework and the Dataset}
\label{sec:release}

\subsection{Framework Release}
\label{sec:framework}

The framework is released as two repositories, split the same way the system
itself is: \code{q-matrix-\discretionary{}{}{}agents}, the code layer (orchestrator, agents, skills,
the FastAPI service, and the dashboard), and \code{q-matrix-\discretionary{}{}{}kb-template}, a
skeleton of the knowledge base's directory structure with no curriculum content.

\code{q-matrix-\discretionary{}{}{}kb-template} ships empty on purpose: directories,
\code{.gitkeep} files, per-directory README files, and one seeded ruleset
(\code{rulesets/\discretionary{}{}{}universal\_rules.md}), plus a fake
\code{EXAMPLE\_BOARD/\discretionary{}{}{}EXAMPLE\_SUBJECT/\discretionary{}{}{}Grade 8} tree demonstrating the expected
shape, which a user deletes before populating the template with their own textbook
PDFs and curriculum documents that they hold the rights to use. Structural changes
to the layout originate in \code{q-matrix-\discretionary{}{}{}agents}'s \code{kb\_access} skill and
land in the template afterward, not the reverse; the template itself is not open to
structural pull requests, curriculum-data contributions belong in
\code{q-matrix-\discretionary{}{}{}dataset} instead. The code layer supports both
Check~2 modes described in \S\ref{sec:verify}, selected with
\code{-{}-check2-mode}. Report-only mode was added in commit \code{bd32d7e} and is
not exercised by any run reported in \S\ref{sec:results}.

\subsection{The Dataset Release}
\label{sec:dataset}

This subsection covers the dataset release specifically. The dataset repository is
data only, no pipeline code, no prompts, no internal run history:
\code{q-matrix-\discretionary{}{}{}dataset} v1.0, \url{https://github.com/MaximusTitan/q-matrix-dataset}.

The release is a static, point-in-time snapshot, not a live mirror of the knowledge
base; subsequent versions would be deliberate re-exports rather than automatic
updates. The canonical artifact is one \code{confirmed\_\discretionary{}{}{}curriculum.csv} per chapter
under a \code{Board/\discretionary{}{}{}Subject/\discretionary{}{}{}Grade/\discretionary{}{}{}Chapter} directory tree that mirrors the source
knowledge base exactly. A derived graph export (\code{graph/graph-core.json},
\code{graph/\discretionary{}{}{}concept-details.json}, \code{graph/meta.json}) is shipped alongside it
as a convenience artifact, mechanically regenerated from those same CSVs by a
deterministic exporter that makes no LLM or network calls. The export is a superset
of the CSVs rather than a view of them: alongside the 6,006 prerequisite edges
asserted in the CSVs it adds 1,442 edges that the exporter infers from
already-judged skill-level relations (\S\ref{sec:viz}), and it attaches per-edge
provenance, a rationale and an asserted-versus-derived flag, that the CSVs do not
carry. For any value present in both, the CSVs are authoritative; for the inferred
edges and the provenance fields there is no CSV counterpart, so a consumer needing
either must read the export.

Textbook PDFs and official board curriculum documents were never part of this
dataset repository to begin with: they live in the framework's knowledge base,
which is not itself released publicly; a \code{kb-template} skeleton (folder
structure only, no textbook or curriculum-document content) is released instead,
for users to populate themselves. Pipeline internals, the prompt library, and the
dated escalation history are out of scope for this repository: they belong to the
framework rather than to the dataset. One clarification on the ruleset, which is
not withheld: the knowledge base holds a single ruleset,
\code{universal\_rules.md}, and \code{q-matrix-\discretionary{}{}{}kb-template} ships a byte-identical
copy, so the criterion Check~1 enforced across every run reported here is already
public.

The following three subsections give a brief exploratory data analysis (EDA) of the
release: coverage by subject, graph-level statistics, and the CSV schema itself.

\subsubsection{Coverage}
\label{sec:coverage}

Coverage is complete within CBSE's Mathematics, Environmental Science, and Science
curricula (Table~\ref{tab:coverage}): every chapter present in the source knowledge
base for these three subjects has reached confirmed status, so there is no
attempted-but-unconfirmed remainder inside this scope. Subjects and boards outside
it (Hindi, English, Social Science, ICT; any non-CBSE board) have no confirmed
chapters at all and are absent rather than filtered.

\begin{table}[htbp]
\centering
\caption{Dataset coverage by subject.}
\label{tab:coverage}
\small
\begin{tabular}{lccc}
\toprule
\textbf{Subject} & \textbf{Grades} & \textbf{Chapters} & \textbf{Concepts} \\
\midrule
Mathematics           & 1--10 & 128 & 1{,}620 \\
Environmental Science & 3--5  & 32  & 571 \\
Science               & 6--10 & 63  & 1{,}578 \\
\midrule
\textbf{Total}        &       & \textbf{223} & \textbf{3{,}769} \\
\bottomrule
\end{tabular}
\end{table}

A concept's graph identity is scoped by subject, grade, and chapter, so the three
per-subject concept counts sum exactly to the deduplicated total with no
cross-subject collision to reconcile, even where a concept name repeats verbatim
across chapters.

\subsubsection{Graph Statistics}
\label{sec:graphstats}

\begin{table}[htbp]
\centering
\caption{Node, edge, and integrity counts for the released graph.}
\label{tab:graphstats}
\small
\begin{tabular}{@{}lr@{}}
\toprule
\textbf{Quantity} & \textbf{Count} \\
\midrule
Chapter CSV files                                  & 223 \\
Concept--skill rows                                & 7{,}145 \\
Concept nodes (deduplicated)                       & 3{,}769 \\
Prerequisite edges, total                          & 7{,}448 \\
\quad L1: within chapter                           & 4{,}272 \\
\quad L2: cross-chapter, same grade and subject    & 612 \\
\quad L3: prior grade, same subject                & 2{,}564 \\
Mechanically derived edges                         & 1{,}442 (19.4\%) \\
Cross-subject edges (via L3)                       & 156 \\
Isolated concepts (no edge in either direction)    & 214 (5.7\%) \\
Cyclic components (concepts involved; largest)     & 88 (210; 7) \\
Unresolved references                              & 0 \\
Self-loops                                         & 0 \\
\bottomrule
\end{tabular}
\end{table}

The 156 cross-subject edges are a byproduct of L3 rather than an implementation of
the unimplemented L4 level (\S\ref{sec:arch}, \S\ref{sec:discussion}): CBSE
introduces Science as a standalone subject only from Grade~6, so a Grade~6 Science
concept's prior-grade prerequisites legitimately resolve into Environmental Science
for Grades~3--5. These edges cross a subject label while remaining within a single
curricular lineage, which is what L3 is defined to traverse.

Every prerequisite reference in the release resolves to an actual concept or skill
within the stated scope; the exporter's integrity check reports zero unresolved
references and zero self-loops. This is a structural guarantee only: it catches
hallucinated targets, not judgments that are wrong but plausible
(\S\ref{sec:discussion}). Two chapter-level warnings are recorded in
\code{graph/meta.json}, where two Mathematics Grade~4 files carry a stale chapter
name in the \code{chapter} column following a folder rename; the directory path is
authoritative in both cases.

\subsubsection{Schema}
\label{sec:schema}

Every \code{confirmed\_\discretionary{}{}{}curriculum.csv} shares the same base header, though not
always the same length: four columns locate the row in the curriculum hierarchy
(\code{board}, \code{subject}, \code{grade}, \code{chapter}); two carry the mapping
itself (\code{concept}, a noun phrase naming one curriculum concept, and
\code{skill}, one verb-led observable capability tied to it); and up to six more
carry prerequisites as JSON arrays encoded as text, at both concept and skill
granularity for L1 and L2 always, and for L3 only where that chapter is eligible
for it. 198 of the 223 released chapters carry all twelve columns; the remaining 25
carry ten, with no L3 columns at all rather than empty ones. These 25 are not
missing L3 by omission: they are the first grade taught in their subject (Maths
Grade~1, 13 chapters; Environmental Science Grade~3, 12 chapters), and L3
prerequisites are defined against an earlier grade in the same subject, so a
subject's first grade has no earlier grade to map against and is structurally
ineligible for L3, not merely unattempted. An empty array denotes no prerequisite
found at that level for that row; a missing column denotes that level does not
apply to that chapter. The CSVs contain no identifiers: references are resolved by
exact name match scoped to the level's stated grade and chapter, and the graph
exporter mints stable IDs downstream. A flag distinguishing directly asserted edges
from mechanically derived ones is preserved for every edge in
\code{graph/\discretionary{}{}{}concept-details.json}, alongside a model rationale whose coverage is
uneven: every L2 and L3 edge carries one, 612 and 2,564 respectively, while only
475 of the 4,272 L1 edges do, or 11\%.

One property of this schema matters enough to state where a data consumer will
encounter it, rather than only in \S\ref{sec:discussion}. The \code{concept} and
\code{skill} columns passed the Check~1 and Check~2 verification loop of
\S\ref{sec:verify}; the six prerequisite columns did not. Prerequisite mapping is
verified for well-formedness only, by the two-attempt parse retry described in
\S\ref{sec:prereq}, so those columns carry no correctness check. That status is
uniform, holding for every prerequisite value in every row of every file, which is
why we document it here rather than encode it as a per-row column that would repeat
one constant 7,145 times.

\subsubsection{Attribute Granularity and Downstream Use}
\label{sec:granularity}

The Q-matrix a consumer builds from this release maps assessment items to skills,
so the released skill rows are the attribute set. That mapping is model-agnostic in
form: any cognitive diagnostic model, and equally any knowledge-tracing approach,
consumes an item-by-attribute matrix. The granularity, however, is not
model-agnostic, and a consumer should check it against their intended model before
use.

The natural unit is the chapter, since a chapter is the scope at which the
curriculum states its objectives and at which an assessment is normally set. Per
chapter the release carries a median of 31 concept-skill pairs, a mean of 32.0, and
a range from 7 to 132. Concepts, if used as attributes instead, run a median of 15
with a range from 3 to 79.

Those counts matter because model families scale differently. Conjunctive and
disjunctive models of the DINA, DINO and G-DINA family partition students across
$2^K$ latent classes for $K$ attributes, which in practice limits $K$ to roughly
twenty or thirty without additional structure. Measured against that limit, 81\% of
released chapters exceed twenty skills and 51\% exceed thirty, so a consumer
fitting one of these models at full skill granularity will need to select a subset
of attributes, aggregate skills within a concept, or impose a hierarchy. Approaches
that instead maintain a separate estimate per attribute, rather than enumerating
joint profiles, scale linearly in $K$ and face no such ceiling; there the binding
constraints are item coverage, since an attribute no item assesses cannot be
estimated at all, and identifiability, since attributes always assessed together by
the same items cannot be separated from one another.

The prerequisite edges bear on the first case. A prerequisite relation renders much
of the joint attribute space inadmissible, because a profile that has mastered a
skill without its prerequisite is not reachable, and attribute-hierarchy approaches
use exactly this to make large $K$ tractable. The L1 edges within a chapter are
therefore usable as hierarchy constraints, subject to the verification caveats in
\S\ref{sec:discussion}.

One scoping limitation applies throughout. Skill identity in this release is
chapter-local. Skills and concepts are resolved by exact name match within the
stated grade and chapter (\S\ref{sec:schema}), and we do not attempt to determine
whether a skill in one chapter denotes the same competency as a similarly worded
skill in another. Exact-string recurrence across chapters is low, 0.4\% of distinct
skill strings and 3.8\% of distinct concept strings, so the release contains almost
no accidental duplication. But the same fact means it provides no mechanism for
recognising a recurring competency across chapters or grades: a consumer tracking a
learner's state across a curriculum, rather than within a single chapter, must
supply their own skill-matching step. Canonicalising skills across chapters is left
to future work.

\subsection{Reproducibility}
\label{sec:repro}

Every agent calls its model through a single wrapper (\code{skills/llm.py}), and
model selection is recorded per agent in every run record from schema v3 onward.
Three identifiers appear across the corpus, logged exactly as follows:
\code{anthropic/\discretionary{}{}{}claude-sonnet-5}, used by the Generator, both Eval checks, Map
Extraction, all three prerequisite levels, and Doctor re-evaluation, observed
2026-07-14 to 2026-07-23; \code{openai/\discretionary{}{}{}gpt-5.4-mini}, used by Doctor, Rules Doctor,
Revision, Judge, Prerequisite L1, by Eval before 2026-07-14, and by the Generator
and Map Extraction in 12 early runs, observed
2026-07-09 to 2026-07-23; and \code{anthropic/\discretionary{}{}{}claude-sonnet-4-6}, used by the
Generator and Map Extraction in the early period with incidental Revision and
Doctor calls, observed 2026-07-09 to 2026-07-11.

These are gateway routing aliases, not provider snapshot identifiers. All calls
route through the Vercel AI Gateway (\url{https://ai-gateway.vercel.sh/v1},
addressed through its OpenAI-compatible interface), which resolves an alias to
whichever provider snapshot is current at request time; the resolved snapshot is
not returned in the response and is not recorded. No dated snapshot identifier is
available for any run reported here, and none can be recovered retrospectively.

Apart from an output-length limit, no generation parameters were set, and none were
recorded. In the 241 full-pipeline runs the wrapper issued each call with
\code{model}, \code{messages} and a fixed \code{max\_tokens} of 8,096, set when the
wrapper was first written (commit \code{e1bb24c}), adding a tool definition and a
forced \code{tool\_choice} for structured calls. Eight Revision calls, all before
the configuration boundary, returned exactly 8,096 output tokens, so those
rewritten prompts were most likely cut off at the limit. The limit was removed in
commit \code{343a1ee}, after the last full-pipeline run; 193 of the 362 per-chapter
L2 and L3 records were produced after its removal. Temperature, top-p, seed, stop sequences, and penalty
parameters are not passed at any call site, are not configurable through the agent
interfaces, and appear nowhere in the run records. Every call otherwise used the
provider's server-side defaults for the resolved snapshot. The pipeline is
consequently not deterministic and was not run under a fixed sampling configuration:
repeated execution of the same chapter is expected to produce different CSVs. The
only repeated-run measurement we hold comes from the ablation
(\S\ref{sec:escalation}): two first attempts by the same model on the same chapter
differ in concept coverage by 0.09 on average. Retry behaviour is the one fixed parameter, at
three attempts five seconds apart, on rate-limit and internal-server errors only.
\code{requirements.txt} pins no versions and the repository carries no lockfile, so
the client SDK version in use during the reported runs is not recorded either.

The full corpus spans 2026-06-26 to 2026-07-24. Structured run records with UTC
timestamps cover 2026-07-08 to 2026-07-23; 26 escalation folders dated earlier
predate the run-record schema and carry no machine-readable timestamp. Of 325
structured records, 241 are full-pipeline runs, the population \S\ref{sec:results}
reports on, alongside 68 L2-only, 14 L3-only, and 2 L1-only prerequisite runs, under
three schema versions (v3: 322, v2: 1, v1: 2).

Taken together, these mean the reported results are not exactly reproducible, and we
state that rather than implying otherwise. A reader seeking to replicate the
pipeline should expect to re-derive a subject's base prompt through the cold-start
path (\S\ref{sec:start}) and to supply their own curriculum sources; what transfers
is the architecture and the ruleset, not the accumulated configuration behind the
numbers in \S\ref{sec:results}.

\subsection{License and Citation}
\label{sec:license}

The framework, \code{q-matrix-\discretionary{}{}{}agents} and \code{q-matrix-\discretionary{}{}{}kb-template}, is released
under the Apache License 2.0. The dataset is released separately under the Creative
Commons Attribution-ShareAlike 4.0 International license (CC BY-SA 4.0), permitting
sharing and adaptation for any purpose, including commercially, with attribution
and under the same license terms for derivative datasets. The dataset's license
covers the derived data in the repository only and does not extend to any textbook
or curriculum-document source material, none of which is included. Users of either
release are asked to cite this paper; users of the dataset specifically are also
asked to cite the dataset release.

\section{Results}
\label{sec:results}

We report results from the system's actual run history rather than a separate
benchmark, since every chapter mapped so far is itself a real invocation of the
full pipeline. Across the escalation logs, run records, and run history retained in
the knowledge base, 687 run-record files exist (the 325 structured records of
\S\ref{sec:repro} plus 362 per-chapter L2 and L3 records stored beside each
chapter's textbook); of these, 241 went through the
full Check~1/Check~2 verification pipeline described in
\S\ref{sec:verify}--\S\ref{sec:repair}, covering 168 distinct chapter identities,
154 of which correspond to chapters in the released dataset (\S\ref{sec:release}).
The other 14 are Grade~4 Mathematics chapters from a superseded NCERT edition, whose
outputs the release does not carry. Conversely, 69 released chapters have no
full-mode run record at all: 56 predate the introduction of run logging, and 13 are
the current-edition Grade~4 Mathematics chapters, all confirmed instead through a
legacy concept-skill-map schema combined with the L2 and/or L3 prerequisite-only
pipeline. A further caveat applies to the statistics themselves: the pipeline's configuration
changed at run 77 (\S\ref{sec:escalation}), so the 241 runs span two pipeline
configurations, and the subsections below report each separately where the
distinction matters. The statistics in this section therefore describe most, but not
all, of the released dataset's confirmation history, and only the core concept-skill
generation loop within it (Generator, Eval, Doctor, Rules Doctor, Revision). All of
these runs use Check~2's gate mode; report-only mode (\S\ref{sec:verify}) is not
evaluated here. Prerequisite mapping (L1, L2, L3) has no pass/fail data to report, since none of the
three levels pass through this verification loop at all (\S\ref{sec:prereq}); its
production statistics are reported separately, as dataset description, in
\S\ref{sec:release}.

\subsection{Pass Rates}
\label{sec:passrates}

Two pass-rate figures are worth reporting, and they answer different questions, so
we keep them distinct. The direct pass rate, the fraction of the 241 full-pipeline
runs in which some Generator output passed both Check~1 and Check~2 without a Doctor
or Rules Doctor patch, was 41.5\%. It is not an unaided rate: a first attempt passed
both checks in only 1 of the 236 runs whose records allow the check
(\S\ref{sec:escalation}; the other 5 hit a Check~2 parse failure on their first
attempt), so nearly every direct pass came on a later attempt, after
at least one round of check feedback. The run-level confirm-without-escalation rate,
the fraction reaching a confirmed state once Doctor, Rules Doctor, and Revision's
repair cycle are counted but before any human involvement, was 67.6\%. The gap
between these two, 63 runs, consists of 61 runs confirmed through a Doctor or Rules
Doctor patch with no direct pass and 2 confirmed after manual corrections to the
concept-skill map (\S\ref{sec:repairmech}). Counted instead by what produced the
confirmed CSV, 96 runs (39.8\%) were confirmed on unpatched Generator output, 65
(27.0\%) on a patch, 51 from Doctor and 14 from Rules Doctor, and 2 on a manual
correction; the direct-pass count exceeds 96 because in 4 runs Judge preferred an
earlier patched candidate over a later direct pass.

Both figures measure agreement between the system's output and criteria the system
itself produced. Check~1 tests compliance with the accumulated ruleset; Check~2
tests coverage against a concept-skill map the pipeline extracted. Neither consults
an expert-authored Q-matrix or any other external reference, so neither figure is
evidence about mapping correctness, and neither is comparable to the validation
statistics (expert alignment, model fit, or prediction accuracy) reported by Zhang
et al. (2026a), Du et al. (2026), Fan, J. et al. (2026), or Fan, K. et al. (2026).

Both figures also pool the two pipeline configurations noted above, and the two
periods behave very differently. Across the 76 runs before the change the direct
pass rate was 7.9\% and the confirm-without-escalation rate 15.8\%; across the 165
runs after it they were 57.0\% and 91.5\%. The pooled figures are averages over a
non-stationary process and describe neither configuration accurately. A reader
wanting the system as it currently stands should take the later figures; a reader
wanting the whole history should take the pooled ones and read
\S\ref{sec:escalation} before interpreting the difference.

Both figures above are run-level, not chapter-level, and the distinction matters: 71
of the 168 distinct chapters behind these 241 runs were resubmitted after an
escalated first run, so a single chapter can contribute to both the escalated and
the eventually-confirmed tallies. At the chapter level, 162 of the 168 distinct
chapters (96.4\%) eventually reached a confirmed state through the pipeline itself,
counting resubmissions; of the remaining 6, 4 were confirmed through a separate,
out-of-band mechanism despite never passing within this tracked pipeline, and 2,
both Grade~4 Mathematics chapters from the superseded edition, were never confirmed. No chapter
required more than 3 runs to resolve. That nearly every chapter was eventually
confirmed needs stating carefully: the attempt loop terminates only on confirmation
or escalation, an escalated chapter is resubmitted until it is confirmed unless it
is dropped (four escalated once and were never resubmitted), and confirmation is
granted by the first author. A near-complete eventual-confirmation figure is therefore
largely entailed by the procedure and carries no information about output quality.

Of the 241 records, 163 resolved organically, without escalation, and 78 escalated.
The distribution of \emph{when} organic resolutions occurred is itself informative.

\begin{table}[htbp]
\centering
\caption{Last attempt reached by each run that was confirmed. The confirmed CSV can
come from an earlier attempt than the one tabulated, and at any attempt can be a
Doctor or Rules Doctor patch rather than the Generator's own output; see the text.}
\label{tab:attempts}
\small
\begin{tabular}{@{}cc@{}}
\toprule
\textbf{Attempt} & \textbf{Runs confirmed} \\
\midrule
1 & 21 \\
2 & 69 \\
3 & 42 \\
4 & 20 \\
5 & 7  \\
6 & 4  \\
\midrule
\textbf{Escalated} & \textbf{78} \\
\bottomrule
\end{tabular}
\end{table}

\begin{figure}[htbp]
\centering
\figplace{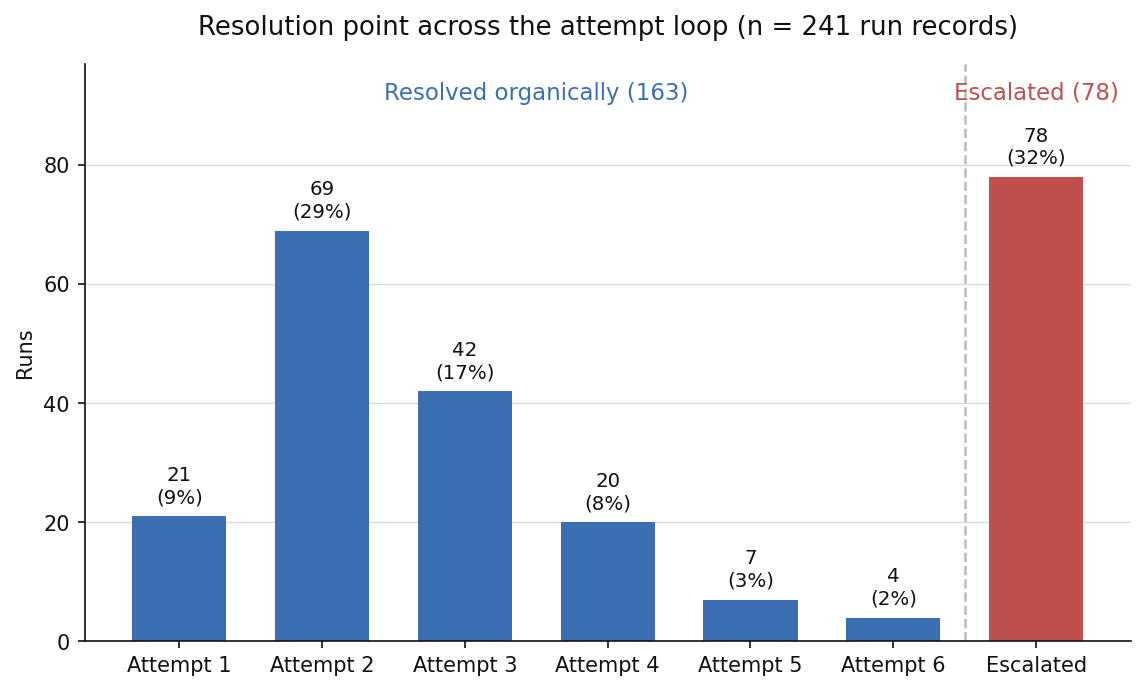}{0.8}
\caption{Distribution of the last attempt reached by runs that were confirmed, with
escalated runs shown separately (see Table~\ref{tab:attempts}).}
\label{fig:attempts}
\end{figure}

Resolution peaks sharply at attempt~2 and tapers toward the ceiling: most runs that
resolve at all do so early, and persistence past attempt~4 yields diminishing
returns. The table pools the two pipeline configurations, whose attempt budgets
differ, so its upper buckets are not comparable across the whole history. No run
before the change reached a fourth attempt, because the budget was three, and all 31
confirmations at attempts 4 through 6 come from the 165 later runs. Reported
separately, the earlier period's confirmations fall at attempts 1, 2 and 3 (1, 3 and
8 runs) and the later period's at attempts 1 through 6 (20, 66, 34, 20, 7 and 4
runs). Two records in this table (one each in the attempt-3 and attempt-4 buckets)
reflect manual concept-skill-map corrections resuming an existing run rather than
organic multi-attempt cycles; excluding them gives 41 and 19 respectively for those
two buckets, with no other change to the table. We report the inclusive figures
above as the primary numbers and note the organic-only variant here for
transparency.

Two further properties of the table need stating. It places each run by the last
attempt it reached, not by the attempt whose CSV was confirmed. The two coincide for
all but 9 runs: 3 early runs in which the loop continued after a candidate had
passed, 4 in which Judge chose an earlier candidate, and the 2 manual corrections
above; in the 7 that are not manual corrections, the confirmed CSV comes from an
earlier attempt than the one
tabulated. And a confirmation at a given attempt need not be the Generator's own
output. Of the 21 runs confirmed at attempt~1, one passed on raw Generator output; in
the other 20 the Generator passed Check~1 but not Check~2, and the Doctor completed
its coverage within the same attempt. This is why the attempt-1 bucket is consistent
with a first attempt passing both checks in only 1 of 236 runs (\S\ref{sec:passrates}).

\subsection{Repair Mechanisms}
\label{sec:repairmech}

Repair itself splits three ways between its mechanisms, for a structural reason
rather than a difference in effectiveness. Doctor and Rules Doctor together
automatically rescued 11.6\% of failed attempts. Revision was invoked on a further
73.9\%, though invocation is not the same as success: 43\% of those attempts
belonged to runs that still escalated despite a revision being attempted. The
remaining 14.5\% (85 of 586) saw no automated repair or revision step logged at all,
and split cleanly into four cases: 78 were a run's final attempt before it
escalated, with no subsequent attempt left to revise toward; 2 were a run's final
attempt despite the run ultimately passing, because an earlier attempt had already
produced the confirmed candidate and the run continued regardless; 3 are
older-schema runs that retried without logging a revision step; and 2 are the manual
human edits excluded above. This reflects \S\ref{sec:repair}'s design, not a
weakness of surgical repair: Doctor and Rules Doctor only ever run on single-check
failures, and most real failures in this data fail both checks at once, routing them
to Revision regardless of how a targeted patch might have performed. (Two
Doctor-schema entries in the underlying records are manual human edits rather than
LLM calls, logged through the same field as automated repairs; these are excluded
from the rescue-rate figure above and from all cost figures reported in this
section.)

Because Check~2's coverage criterion is textbook-derived (\S\ref{sec:start}) while
the Generator works from curriculum documents, it is worth reporting what the repair
loop actually moved between the two. Across the 151 chapters with both a Check-2
demand and a confirmed CSV, Check~2 named 1,591 distinct concepts as missing, and
1,300 of them (81.7\%) appear in the chapter's final confirmed output. In the other
direction it flagged 1,685 concepts as covering nothing expected, of which 1,148
(68.1\%) do not appear in the final output, despite the conservative retention
instruction. Taken together, 1,300 of the 2,576 concepts in those chapters' final
CSVs, or 50.5\%, are traceable to a Check-2 demand rather than to the Generator's
own reading of the curriculum documents. That figure bounds the textbook-driven
share of the released concepts; it does not measure error, since a demanded concept
may be both a genuine curriculum objective and one the Generator initially missed.
It is also the measured cost of gate mode: the share of the output that the
automated resolution path writes in place of the per-chapter judgment report-only
mode leaves to a human.

Coverage repair is also less reliable than the rescue rate alone suggests. Of 271
coverage-Doctor invocations, 140 (51.7\%) produced a candidate that passed Check~2
on re-evaluation, and 85 (31.4\%) were flagged as having regressed previously matched
coverage while patching the gaps. Adding the demanded items and preserving existing
matches are in tension, which is the oscillation between the two checks that
\S\ref{sec:repair}'s routing grid describes.

\subsection{Escalation Rate and the Configuration Change}
\label{sec:escalation}

Escalation rate fell sharply within the available data: 84.2\% (64 of the 76 runs
before the change described below) versus 8.5\% (14 of the 165 after it). We take
the breakpoint at run 77 because that is where the pipeline's configuration changed.
It also sits inside a wide band of equivalent splits: the last escalation of the
earlier period is run 75 and the next is run 98, so any breakpoint between them
yields the same counts of 64 and 14.

We report the fall as an observed change in the system's behaviour, not as evidence
of anything the system learned on its own. The boundary is a single commit
(\code{84151f7}), together with a knowledge-base commit made the same day, and
between them they change nine things at once. On the judging side, the Check~1 gate
was rewritten from one LLM judgment over the full 31-rule ruleset to deterministic
code for structural rules plus an LLM judging eight blocking content rules; a
non-blocking advisory flag channel was added, and 66\% of Check~1 passes after the
change carry at least one such flag; and Eval, which runs both checks, moved from
gpt-5.4-mini to claude-sonnet-5, so the escalation criterion was itself applied by a
different model on each side of the boundary. On the generating side, the Generator
and Map Extraction moved to claude-sonnet-5 from claude-sonnet-4-6 (gpt-5.4-mini in 12
early runs), a retry now edits
the previous attempt's CSV rather than regenerating from scratch, and the Doctor and
Rules Doctor prompts were revised. In the loop, the attempt budget rose from three to
six and the plateau stop of \S\ref{sec:repair} was introduced. In the prompts, the
Revision agent's rewrites, which until then had been written back to the shared
prompt library and carried into later chapters, stopped being persisted; the
Mathematics base prompt those rewrites had built up was pruned by hand from 1,544 to
356 words, and five grade-level overrides, also written by the pipeline, were
deleted.

The budget change alone accounts for part of the fall. Every one of the 64
escalations in the earlier period was a budget exhaustion at attempt~3. Recomputing
the later period under a matched three-attempt cap, counting as escalated every run
that had produced no passing candidate by its third attempt, gives an escalation
rate of 26.1\% (43 of 165) rather than 8.5\%, so roughly a quarter of the observed
drop (17.6 of its 75.7 percentage points) is attributable to permitting more
attempts rather than to any improvement in what an attempt produces. This measures
the budget's effect under the later configuration; the earlier configuration never
ran with six attempts, so the effect there is not observed.
First attempts did not improve either. Their mean concept coverage against the map
fell from 0.40 before the boundary to 0.31 after it, and a first attempt passed both
checks in 1 of 76 runs before and 0 of 160 after, so the fall came from how the loop
converges rather than from better first drafts.

To test the explanation most readily assumed, the model upgrade, we ran a scoped
ablation after the corpus closed. Twelve Mathematics chapters, drawn from those that
escalated before the boundary and spread across Grades~3 to~10, were re-run with the
Generator on claude-sonnet-4-6 or claude-sonnet-5 while everything else was held
fixed: Eval on claude-sonnet-5, each chapter's existing concept-skill map, and the
post-boundary pipeline. Because retry feedback carries Check~2's missing items to the
Generator by name (\S\ref{sec:verify}), the primary measures are taken on the first
attempt only, concept coverage against the map and Check~1 pass or fail, with two
replicates per model and chapter. On neither did the upgraded model do better. Its
first-attempt coverage differed from the older model's by a mean of $-0.02$ (median
0.00; higher in 2 chapters, lower in 5, tied in 5), less than the 0.09 mean
difference between two runs of the same model on the same chapter. Its first attempts
passed Check~1 in 12 of 24 runs against 19 of 24 for claude-sonnet-4-6, a difference
in the older model's favour but not a significant one: by chapter, claude-sonnet-4-6
did better in 4, claude-sonnet-5 in none, and 8 were tied, giving $p = 0.125$ on a
two-sided sign test, the smallest value four untied chapters allow. On six of the
chapters we also ran the full loop at both budgets. Each model passed 4 of 6 at three
attempts and 5 of 6 at six, the larger budget rescuing one chapter for each model and
losing none, and 9 of the 18 passes came through a Doctor patch rather than directly
from the Generator. The ablation cost \$12.90 in API spend.

Three conclusions follow. The Generator's model upgrade does not account for the
fall. The budget accounts for part of it, consistent with the estimate above, though
six chapters are too few to refine that estimate. What remains lies on the judging
side and in the loop's mechanics: the Check~1 rewrite, which cut the share of
Generator evaluations failing Check~1 from 77.6\% to 30.7\%, the change of Eval
model, and the retry and prompt changes listed above. The ablation varies none of
these, and we name them as untested candidates rather than findings. Two explanations
can be set aside. Rule accumulation did not occur: \code{universal\_rules.md} is
unchanged from a month before the first reported run, no grade-scoped ruleset was
ever created, and the boundary removed the one mechanism that had carried prompt
changes across chapters, so the fall coincides with the system retaining less across
runs, not more. Nor is the fall compositional: restricting to Mathematics alone
reproduces it (84.2\% to 7.2\%), it holds within every grade, and among the 65
chapters run in both periods, 64 escalated before the change while 3 of 66 runs
escalated after it. Rule-violation counts show no trend within either period.

The ablation's limits are specific. It covers Mathematics only, because no Science or
Environmental Science chapter ran before the boundary, and twelve chapters, with six
in the full-loop runs, two of which were chosen because they had escalated after the
boundary. It ran on code later than the corpus, in which Check~2 returns structured
output and no output-token limit is set (\S\ref{sec:repro}). Model identifiers are
gateway aliases, so the claude-sonnet-4-6 it used need not be the snapshot that ran
in July. Full-loop retries converge on a target shown to the Generator, so they
measure repair under feedback rather than unaided generation. And two runs failed on
a malformed Check~2 reply and were rerun.

\subsection{Cost}
\label{sec:cost}

That same model transition, however, does yield a legitimate and separable
comparison: cost across the three models used at different points in the
system's history, gpt-5.4-mini, claude-sonnet-4-6, and claude-sonnet-5, abbreviated
in Table~\ref{tab:cost} below as gpt-5.4-mini, sonnet-4-6, and sonnet-5. Costs here
are per run, not per call and not per chapter: an important distinction, since
Generator and Eval are invoked once per attempt, and a run needing several attempts
triggers several calls to the same agent. Sample sizes accordingly count run records
rather than chapters, which is why a figure such as Revision's $n{=}210$ exceeds the
number of distinct chapters without inconsistency. Per-run cost is also
right-skewed, because it scales with attempt count: the Generator's sonnet-5 mean of
\$0.1299 sits against a median of \$0.0979, and Eval's sonnet-5 mean of \$0.3534
against a median of \$0.2846.

\begin{table}[htbp]
\caption{Mean cost per full-pipeline run (USD) by agent and model, with standard deviation and sample size. $n$ is the number of run records in which that agent-model pair fired: because an agent may be invoked several times within a run, each figure aggregates all of that agent's calls in that run rather than reporting a per-call cost. Cells with $n < 5$ are not reported. Dashes indicate the pair never occurred. Eval includes the re-evaluation calls issued after each Doctor and Rules Doctor repair. Prerequisite (L1) is the same underlying spend reported per chapter later in this section, where it is blended across models and so differs from the per-model figures here; L2 and L3 run outside full-mode runs and are reported only there.}
\label{tab:cost}
\footnotesize
\begin{tabular}{@{}lccc@{}}
\toprule
\textbf{Agent} & \textbf{sonnet-4-6} & \textbf{sonnet-5} & \textbf{gpt-5.4-mini} \\
\midrule
Generator     & \$0.1128 (SD 0.0236, $n{=}61$)  & \$0.1299 (SD 0.0974, $n{=}165$) & \$0.0269 (SD 0.0135, $n{=}12$)  \\
Eval          & $n{=}2$, not reported           & \$0.3534 (SD 0.2318, $n{=}165$) & \$0.0470 (SD 0.0245, $n{=}67$)  \\
Doctor        & $n{=}2$, not reported           & ---                             & \$0.0169 (SD 0.0094, $n{=}171$) \\
Rules Doctor  & $n{=}1$, not reported           & ---                             & \$0.0119 (SD 0.0067, $n{=}118$) \\
Revision      & \$0.2389 (SD 0.0855, $n{=}5$)   & ---                             & \$0.0147 (SD 0.0092, $n{=}210$) \\
Judge         & ---                             & ---                             & \$0.0071 (SD 0.0013, $n{=}5$)   \\
Map Extraction& \$0.0292 (SD 0.0107, $n{=}57$)  & \$0.0349 (SD 0.0127, $n{=}64$)  & \$0.0077 (SD 0.0013, $n{=}12$)  \\
Prereq.\ (L1) & $n{=}1$, not reported           & \$0.1097 (SD 0.0389, $n{=}8$)   & \$0.0083 (SD 0.0036, $n{=}153$) \\
\bottomrule
\end{tabular}
\end{table}

\begin{figure}[htbp]
\centering
\figplace{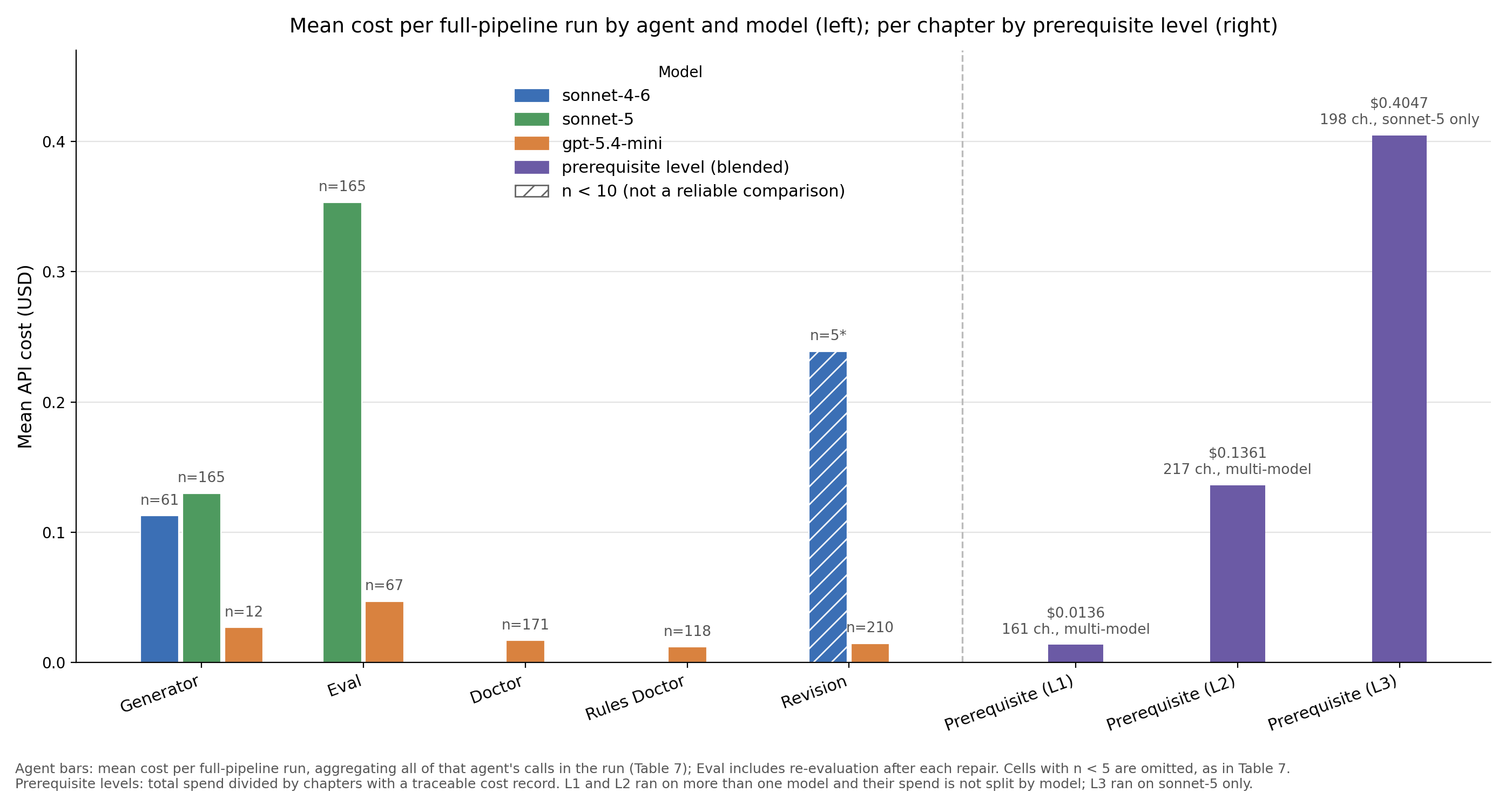}{0.85}
\caption{Mean API cost by agent and model, per full-pipeline run, and by prerequisite level, per chapter (values as in Table~\ref{tab:cost}).}
\label{fig:cost}
\end{figure}

Prerequisite mapping's cost, among chapters with a traceable run record: L1 \$0.0136
(161 chapters, \$2.19 total), L2 \$0.1361 (217 chapters, \$29.52 total), L3 \$0.4047
(198 chapters, \$80.14 total, confirmed sonnet-5-only). L1's spend is already
contained within the full-mode run records, because L1 runs automatically inside the
same run rather than as a separate invocation, so it must not be added to the
core-loop total a second time. These populations are narrower than ``every chapter
that has this level mapped'': L1 and L2 columns are present in all 223 released
chapters, but only 161 and 217 of them respectively have a traceable cost record
behind that data, the rest were confirmed through the out-of-band pathway described
in \S\ref{sec:results}'s introduction, with no logged run to attribute a cost to.
L3's columns, by contrast, are present only where L3 was actually eligible and
attempted (198 of 223, \S\ref{sec:schema}), so 198 already is the full attempted
population. L3 is the single most expensive per-chapter figure anywhere in this data,
roughly 1.15 times Eval's per-run mean on sonnet-5, the highest figure in
Table~\ref{tab:cost}, plausibly reflecting the larger
cross-grade context L3 must reason over, though this is not confirmed against call
counts. The per-level model attribution is as follows. L1 ran on gpt-5.4-mini in 153
of the 162 full-mode runs that produced it, with 8 on claude-sonnet-5 and 1 on
claude-sonnet-4-6. L2 ran on claude-sonnet-5 in 230 of its 232 records and on
gpt-5.4-mini in 2. L3 ran on claude-sonnet-5 throughout its 212 records. The
per-chapter cost figures above remain blended for L1 and L2 because the records do
not separate spend by model at those levels, but the attribution itself is no longer
uncertain, and it is worth stating plainly that the level contributing the most edges
is also the one produced by the cheaper model, for the context-breadth reason given
in \S\ref{sec:prereq}.

gpt-5.4-mini is consistently the cheapest option across every agent, by a wide
margin. Restricting the comparison to cells with $n \geq 5$ on both sides, it runs
3.8 to 4.8 times cheaper for Map Extraction and the Generator, 7.5 times cheaper for
Eval, and 13.2 and 16.3 times cheaper for Prerequisite L1 and Revision respectively.
The upper endpoint rests on Revision's sonnet-4-6 cell, whose $n{=}5$ is the
smallest retained in the table, so it is indicative only; without it the span is
3.8 to 13.2. Read alongside the escalation-rate finding above, this points to
a real cost-quality tradeoff rather than a simple ``cheaper is worse'' story:
gpt-5.4-mini's low per-call cost as Eval's model coincided with a much higher
escalation rate, while sonnet-5's higher per-call cost coincided with a much lower
one, though, as stated above, these are confounded with time and cannot be cleanly
separated as cause and effect. Doctor, Rules Doctor, and Revision never ran on
sonnet-5 at all. Their sonnet-4-6 samples are too small to support a mean: Doctor's
$n{=}2$ and Rules Doctor's $n{=}1$ are suppressed in Table~\ref{tab:cost} rather than
reported, and Revision's $n{=}5$ is retained only alongside its standard deviation.
The core generation loop averaged \$0.444 per run across the 241 full-pipeline runs,
or \$0.64 per distinct chapter (168), since some chapters required more than one run.
Total observed spend across all 687 run records is \$216.77. Two per-chapter figures
follow from it, and they answer different questions. A chapter processed end to end
through the current pipeline, meaning one that has a full-mode run record and also
appears in the release, cost a mean of \$1.1924 and a median of \$1.1551
($n{=}154$); this is the figure to use when estimating what producing a chapter
costs. Averaged instead across all 223 released chapters, spend is \$0.9721, but that
average is diluted in both directions: 69 released chapters have no full-mode record
and so contributed no core-loop spend, while the numerator includes \$7.22 spent on
14 chapters that are not in the release at all.

All figures in this section are API spend only. They exclude the review time behind
every confirmation, the human handling of 78 escalations and 72
resubmissions across 71 chapters, and the engineering time to build and operate the pipeline.
Wall-clock time is not tracked anywhere in the current run-record schema, so we do
not report or estimate it, and we therefore offer no total cost of ownership and no
comparison against a manual baseline.

\section{Discussion and Limitations}
\label{sec:discussion}

The results in \S\ref{sec:results} measure internal consistency rather than
correctness. Check~1 tests compliance with a ruleset the system authored itself;
Check~2 tests coverage against a concept-skill map that Map Extraction produced and
that is not independently validated. Neither consults an external standard, and for
most of the run history the same model served as both Generator and Eval, so
self-preference effects cannot be excluded in general. The ablation of
\S\ref{sec:escalation} offers one data point against them: with Eval on
claude-sonnet-5, claude-sonnet-4-6's first attempts passed Check~1 at least as often
as claude-sonnet-5's in every chapter, the opposite of what self-preference would
predict, though the difference is not significant. We therefore report no agreement
statistic against an expert-authored Q-matrix, and claim no psychometric validity
for the released data. Most consequentially, we never fit a CDM: the paper's premise
is that the Q-matrix blocks cognitive diagnosis, but we hold no student response data
and report no model fit, leaving the released mapping a candidate structure of
untested diagnostic utility. Nor do we establish that the released granularity suits
any particular model: \S\ref{sec:granularity} reports the per-chapter attribute
counts and what they imply for different model families, but which granularity is
diagnostically appropriate is itself an empirical question we have not addressed.
The efficiency argument is likewise asserted rather than measured: \$1.19 per chapter
counts API spend only, excludes all human review time, and has no manual baseline
beside it. Two further threats bound the results. The escalation rate's fall from 84.2\% to
8.5\% coincides with a single configuration boundary that changed the verification
gate, the models, the attempt budget and the retry loop together
(\S\ref{sec:escalation}); a scoped ablation rules out the Generator's model upgrade
as the cause and is consistent with a partial budget effect, but leaves the rest
unattributed, so no part of the fall can be credited to the system having learned
anything. And we run no architectural ablation, so nothing establishes that eleven
single-responsibility agents outperform one well-prompted call, though the choice
follows a broader pattern of decomposing educational and knowledge-graph tasks into
specialized agents rather than one generalist call (Santos et al., 2026; Zhao et al.,
2025b).

Within the released artifact, the verification guarantee covers concept-skill mapping
only. None of L1, L2, or L3 passes through the Check~1/Check~2/repair loop: each
issues a single LLM call and writes straight into the confirmed CSV, so all 7,448
prerequisite edges lack that guarantee, and 1,442 of them (19.4\%) are not
model-judged at all but inferred in code from skill-level relations. On the one
quality signal available to us, however, those inferred edges behave better than the
judged ones: they are 19.4\% of all edges but only 2.8\% of the edges caught in the
contradictory pairs described below, so the deterministic inference is not the weaker
half of the graph. The only guardrail is name resolution, which catches fabricated
targets but not edges pointing the wrong way or omitting real dependencies. The graph
is also not a strict DAG, and we can be specific about the reason. 99 concept pairs
carry prerequisite edges asserted in both directions, 198 edges or 2.7\% of the
total, with 194 of those at L1 and 97 of the 99 pairs falling within a single
chapter. Since A and B cannot each precede the other, every such pair contains either
a wrong edge or a relation that does not apply, and inspection finds both. Some pairs
are genuinely co-defining, so that no linear order is more correct, as with
``repeated addition as a basis for multiplication'' and ``multiplication as equal
groups''. Others invert a general concept and its specialisation, as when
``perimeter of a rectangle'' is asserted as a prerequisite of ``perimeter''. With 31
further mutual pairs absorbed into longer loops, these form 88 strongly connected
components over 210 concepts, 68 of them a single mutual pair and 20 spanning three
to seven concepts; we inspected all 20 and found the same mixture, with the general-
and-specific inversion the most frequent identifiable error. This departs from
K12-KGraph's (Liang et al., 2026) validated acyclicity and leaves consumers needing a
topological order to supply their own cycle-breaking policy. It also shows what the
absent correctness check (\S\ref{sec:prereq}) would have caught: inversions of this
kind follow a small number of patterns and are mechanically detectable, and we leave
that check to future work. A further 214 concepts (5.7\%) are isolated, plausibly
standalone in some cases and reflecting mapping gaps in others, which we cannot
distinguish from the data alone.

A separate limitation concerns the coverage criterion in gate mode, the mode behind
every reported result. There, Check~2 measures the candidate against a
textbook-derived map that is written once and never revised (\S\ref{sec:start}), and
coverage of every expected item is mandatory (\S\ref{sec:verify}). An item the
extraction should not have produced is therefore mandatory to include, and no
automated path can remove it: the only remedy within gate mode is editing the map by
hand, which happened twice across 241 runs. \S\ref{sec:repairmech} quantifies the
consequence, that half the concepts in the affected chapters' final output entered
through a Check-2 demand. Report-only mode avoids this by leaving each item to a
human, at the price of the per-chapter review gate mode was built to remove, and we
have not measured how that review performs in practice. For the automated path, the
fix we would build is to stratify the map at extraction time rather than to make it mutable,
tagging each extracted item by where in the chapter it came from, with the gate
requiring coverage of main-exposition items only and treating activity-box and
enrichment-box items as advisory. That keeps the textbook independent of the
Generator's source, which making the map answerable to the curriculum documents would
not, and it avoids a loop in which two mutable artefacts converge by deleting
whatever they disagree about. We leave it to future work.

Three claims need narrowing. Coverage is one board and three subjects: Mathematics
(Grades 1--10), Environmental Science (Grades 3--5), and Science (Grades 6--10) under
CBSE, and these are the subjects with the most explicit prerequisite hierarchy, so
the unmapped ones are the harder cases rather than merely remaining work; L4 is
designed but unimplemented, making L3 the ceiling. Cross-run persistence of the
Revision agent's prompt rewrites is not active in the released codebase. It was
active for the first 76 reported runs, where it carried chapter-specific content into
subject-wide prompts, and was removed at the configuration boundary, after which the
Mathematics seed was pruned by hand (\S\ref{sec:escalation}). The one durable
learning channel that remains, human rejection, was never exercised:
\code{universal\_rules.md} is unchanged from a month before the first reported run,
and no grade-scoped ruleset was ever created. The system therefore accumulated no
rules over this run history, and the prompt persistence it did perform was undone
rather than kept, which is considerably narrower than ``self-tuning'' implies; the
mechanism's effect on output quality is untested rather than merely bounded by review
effort. The review protocol was narrower than ``human verification''
suggests, and we set it out here in full. A single reviewer, the first author, performed
every confirmation. There was no second rater, so no inter-rater agreement is
reported and none is obtainable for this dataset as it stands. The reviewer's
standard is expressed in \code{universal\_rules.md}, the one manually authored
artifact in the system (\S\ref{sec:start}), rather than in a separate written rubric.
No chapter was rejected through the rejection path of \S\ref{sec:selection}, so the
rejection volume is zero. Of the 72 chapters that received more than one run, 62
escalated before the configuration change described in \S\ref{sec:escalation} and
were confirmed after it (61 on their first run under the new configuration, one on
its second), meaning the intervening action in those cases was the configuration
change rather than chapter-specific review; 7 escalated under the stable
configuration and were confirmed on resubmission; 2 escalated in both periods and
never reached a confirmed state within the pipeline; and one, confirmed before the
change, was re-run and confirmed again. Four further chapters escalated on their
only run and were not resubmitted, so 6 of the 168 chapters never reached a
confirmed state within the pipeline at all. For that small residue of genuinely failed runs, the escalation
reports were analysed and the feedback seeding the next run was composed by an LLM
coding agent at the first author's direction rather than by the first author directly, so the
role the design reserves for human judgment was filled, in precisely the cases that
reached it, by another model. None of this is recoverable from the run records, which
do not persist feedback text (\S\ref{sec:kb}); it is reported from the first author's own
account of the process. Taken together, this paper does not evidence that the
system's output is trustworthy enough to hand off without review.

\section{Conclusion}
\label{sec:conclusion}

The barrier to deploying cognitive diagnosis at scale is the Q-matrix, and the
barrier to the Q-matrix is the volume of expert judgment its construction demands.
Curriculum Brain addresses the first half of that judgment rather than eliminating
it: the curriculum's skill space is built once, from official source documents, by a
pipeline of single-responsibility agents that generate, check, and repair their own
output, with human effort reserved for cases the system cannot resolve. Mapping
items against that space, which is the step that would produce a Q-matrix, is not
implemented here. Across 241 verified chapter runs spanning 168 distinct chapters,
166 were eventually confirmed, 162 (96.4\%) through the pipeline itself and 4 out
of band, with the other two superseded-edition chapters outside the release, at an API cost of \$1.19 per chapter processed end to end
through the pipeline, or \$0.97 averaged across the released dataset, and we release
both the framework and the resulting curriculum graph. Confirmation here means the
attempt loop's termination condition together with the first author's acceptance decision,
measured against the system's own criteria rather than any external standard, so it
describes the pipeline's behaviour and not the mappings' correctness. All reported
figures describe Check~2's automated gate mode; the report-only mode is released but
unevaluated.

Two boundaries on this result deserve emphasis rather than burial. Most
consequentially, we never fit a cognitive diagnostic model against student response
data, so the released mapping's diagnostic utility, the very thing this paper's
premise assumes, remains untested; validating it against expert judgment and real
response data is accordingly the most valuable next step, and is the subject of work
now in progress using this framework's output. That work covers the
two stages this paper leaves open: mapping assessment items to the released skills,
and fitting a diagnostic model to the result. The released graph already grounds an
assessment-generation tool in which a user selects skills from a chapter and receives
items targeting them; we note that use as motivation for the artifact rather than as
a result, since we report no measurement of it here. The second is the
escalation-rate fall reported in \S\ref{sec:escalation}: it coincides with a single configuration boundary that changed the verification gate,
the models, the attempt budget and the retry loop at once. The budget change accounts
for roughly a quarter of the fall, a scoped ablation finds no contribution from the
Generator's model upgrade, and the remaining changes, chiefly on the judging side,
are not separated here, so no part of the improvement can be credited to the system
having learned anything. Two narrower
architectural boundaries also matter: the verification guarantee covers concept-skill
mapping only, not the prerequisite graph, which is produced by unverified judgment
rather than the Check~1/Check~2 loop; and the system's durable learning channel,
human rejection, went unused across this run history, so no accumulated rules stand
behind the results reported here and automatic prompt rewriting remains scoped to a
single chapter's attempt loop. Closing these two follows from validation, alongside
extending coverage to the remaining CBSE subjects, to cross-subject prerequisites
(L4), and to additional examination boards.

\section*{References}
\addcontentsline{toc}{section}{References}

\small

\bibent{Bommarito II, M. J. (2025). OpenGloss: A Synthetic Encyclopedic Dictionary
and Semantic Knowledge Graph. \emph{arXiv:2511.18622}.}

\bibent{Chu, Z., Wang, S., Xie, J., et al. (2025). LLM Agents for Education:
Advances and Applications. \emph{arXiv:2503.11733}.}

\bibent{de la Torre, J. (2011). The generalized DINA model framework.
\emph{Psychometrika}, 76(2), 179--199.
\url{https://doi.org/10.1007/s11336-011-9207-7}}

\bibent{Doignon, J.-P., \& Falmagne, J.-C. (1985). Spaces for the assessment of
knowledge. \emph{International Journal of Man-Machine Studies}, 23(2), 175--196.
\url{https://doi.org/10.1016/S0020-7373(85)80031-6}}

\bibent{Du, W., Shen, J., \& Ma, X. (2026). Applying GenAI to optimize Q-matrix
construction for cognitive diagnostic assessment in EFL reading. \emph{Journal of
Intelligence}, 14(5), 79.
\url{https://doi.org/10.3390/jintelligence14050079}}

\bibent{Fan, J., Zhang, Z., \& Du, Y. (2026). From text to structure: Precise
cognitive diagnosis via semantic enhancement and dynamic Q-matrix calibration.
\emph{Applied Sciences}, 16(7), 3477. \url{https://doi.org/10.3390/app16073477}}

\bibent{Fan, K., Bialo, J. A., \& Li, H. (2026). The Use of AI Tools to Develop
and Validate Q-Matrices. \emph{arXiv:2602.08796}.}

\bibent{Fernandes, R. C., Fehring, L., Eimer, T., Lindauer, M., \& Feurer, M.
(2026). Environment-Grounded Automated Prompt Optimization for LLM Game Agents.
\emph{arXiv:2606.17838}.}

\bibent{Junker, B. W., \& Sijtsma, K. (2001). Cognitive assessment models with few
assumptions, and connections with nonparametric item response theory. \emph{Applied
Psychological Measurement}, 25(3), 258--272.}

\bibent{Kim, Y., Lee, Y., Lee, H., \& Lim, S. (2022). Alignment of concepts of
meiosis among curriculum, textbooks, classroom teaching and assessment in upper
secondary school in Republic of Korea. \emph{Journal of Baltic Science Education},
21(2), 232--244.}

\bibent{Liang, H., Lin, Q., Han, Z., Ma, X., Wong, Z. H., Qiang, M., Sun, L., \&
Zhang, W. (2026). K12-KGraph: A Curriculum-Aligned Knowledge Graph for Benchmarking
and Training Educational LLMs. \emph{arXiv:2605.09635} [cs.CL].}

\bibent{Madaan, A., Tandon, N., Gupta, P., et al. (2023). Self-Refine: Iterative
Refinement with Self-Feedback. In \emph{Advances in Neural Information Processing
Systems 36 (NeurIPS 2023)}. arXiv:2303.17651.}

\bibent{Pei, Z., Zhen, H.-L., Kai, S., Pan, S. J., Wang, Y., Yuan, M., \& Yu, B.
(2025). SCOPE: Prompt Evolution for Enhancing Agent Effectiveness.
\emph{arXiv:2512.15374}.}

\bibent{Pryzant, R., Iter, D., Li, J., Lee, Y. T., Zhu, C., \& Zeng, M. (2023).
Automatic Prompt Optimization with ``Gradient Descent'' and Beam Search.
\emph{arXiv:2305.03495}.}

\bibent{Santos, G., Julia, R., \& Nascimento, M. (2026). Agentic Knowledge Tracing:
A Multi-Agent LLM Architecture for Stealth Assessment of Financial Literacy in
Serious Games. Accepted at the IEEE Conference on Games (CoG 2026).
\emph{arXiv:2606.25358}.}

\bibent{Tao, J., Zhao, W., Zhang, Y., Guo, Q., Min, B., Xu, X., \& Liu, F. (2024).
Cognitive diagnostic assessment: A Q-matrix constraint-based neural network
method. \emph{Behavior Research Methods}, 56(7), 6981--7004.
\url{https://doi.org/10.3758/s13428-024-02404-5}}

\bibent{Tatsuoka, K. K. (1983). Rule space: An approach for dealing with
misconceptions based on item response theory. \emph{Journal of Educational
Measurement}, 20(4), 345--354.}

\bibent{Templin, J. L., \& Henson, R. A. (2006). Measurement of psychological
disorders using cognitive diagnosis models. \emph{Psychological Methods}, 11(3),
287--305.}

\bibent{Yang, C., Wang, X., Lu, Y., Liu, H., Le, Q. V., Zhou, D., \& Chen, X.
(2023). Large Language Models as Optimizers. \emph{arXiv:2309.03409}.}

\bibent{Yu, J., Li, C., \& Li, G. (2022). Alignment between biology curriculum
standards and five textbook editions: a content analysis. \emph{International
Journal of Science Education}, 44(14), 2241--2260.
\url{https://doi.org/10.1080/09500693.2022.2119621}}

\bibent{Zhang, Y., Cheng, N., Gao, Y., Li, H., Shi, L., Young, N., Yuan, G., \&
Zhai, X. (2026a). A Framework for Human-AI Q-Matrix Refinement: A NeuralCDM
Evaluation. In \emph{Artificial Intelligence in Education (AIED 2026)}, Lecture
Notes in Computer Science, vol.~16583, 271--278. Springer.
\url{https://doi.org/10.1007/978-3-032-29760-0_30}. arXiv:2604.16398.}

\bibent{Zhang, X., Gong, X., Qin, J., Wang, Q., Liao, J., Wang, Z., Feng, D., \&
Ding, B. (2026b). Beyond Scores: Diagnostic LLM Evaluation via Fine-Grained
Abilities. \emph{arXiv:2604.12191}.}

\bibent{Zhao, X., Wang, G., \& Lu, Y. (2025a). MDKAG: Retrieval-augmented
educational QA powered by a multimodal disciplinary knowledge graph. \emph{Applied
Sciences}, 15(16), 9095.}

\bibent{Zhao, X., Blum, M., Gao, F., et al. (2025b). AGENTiGraph: A Multi-Agent
Knowledge Graph Framework for Interactive, Domain-Specific LLM Chatbots. In
\emph{Proceedings of the 34th ACM International Conference on Information and
Knowledge Management (CIKM '25)}, 6757--6761.
\url{https://doi.org/10.1145/3746252.3761459}}

\bibent{Zhou, Y., Muresanu, A. I., Han, Z., Paster, K., Pitis, S., Chan, H., \& Ba,
J. (2022). Large Language Models Are Human-Level Prompt Engineers.
\emph{arXiv:2211.01910}.}

\bibent{Zhu, Z., Zhong, R., \& Du, L. (2026). A synergistic framework for cognitive
diagnosis with LLM-empowered data augmentation. In \emph{Proceedings of the 16th
International Learning Analytics and Knowledge Conference (LAK26)}, 558--567.
\url{https://doi.org/10.1145/3785022.3785112}}

\end{document}